\documentclass[12pt,preprint]{aastex}

\newcommand{\eu}{SNLS~06D4eu}
\newcommand{\sbv}{SNLS~07D2bv}

\newcommand{\sneia}{SNe~Ia}
\newcommand{\snia}{SN~Ia}
\newcommand{\rindex}{$R_{23}$}
\newcommand{\oindex}{$O_{32}$}

\newcommand{\Msun}{${\rm M}_\odot$}          
 
\newcommand{\Ni}{$^{56}\rm Ni$} 
\newcommand{\Co}{$^{56}\rm Co$} 
\newcommand{\Fe}{$^{56}\rm Fe$} 

\newcommand{\kms}{km s$^{-1}$}

\newcommand{\gp}{$g$\arcmin}
\newcommand{\rp}{$r$\arcmin}
\newcommand{\ip}{$i$\arcmin}
\newcommand{\zp}{$z$\arcmin}
\shorttitle{SNLS~06D4eu}
\shortauthors{SNLS et al.}

\begin{document}

\title{Two superluminous supernovae from the early universe discovered by the Supernova Legacy Survey}


\author{D.~A.~Howell\altaffilmark{1,2},
D.~Kasen\altaffilmark{3,4},
C.~Lidman\altaffilmark{5},
M.~Sullivan\altaffilmark{6},
A.~Conley\altaffilmark{7},
P.~Astier\altaffilmark{8}, 
C.~Balland\altaffilmark{8,9},
R.~G.~Carlberg\altaffilmark{10},
D.~Fouchez\altaffilmark{11},
J.~Guy\altaffilmark{8}, 
D.~Hardin\altaffilmark{8}, 
R.~Pain\altaffilmark{8}, 
N.~Palanque-Delabrouille\altaffilmark{12}, 
K.~Perrett\altaffilmark{13}, 
C.~J.~Pritchet\altaffilmark{14},
N.~Regnault\altaffilmark{8},
J.~Rich\altaffilmark{12},
V.~Ruhlmann-Kleider\altaffilmark{12}
}

\altaffiltext{1}{Las Cumbres Observatory Global Telescope Network,
  6740 Cortona Dr., Suite 102, Goleta, CA 93117}
\altaffiltext{2}{Department of Physics, University of California,
  Santa Barbara, Broida Hall, Mail Code 9530, Santa Barbara, CA 93106-9530}
\altaffiltext{3}{Departments of Physics and Astronomy, University of California, Berkeley, Berkeley, CA 94720-7300 USA}
\altaffiltext{4}{Nuclear Science Division, Lawrence Berkeley National Laboratory, 1 Cyclotron Road, Berkeley CA 94720 USA}
\altaffiltext{5}{Australian Astronomical Observatory, P.O. Box 915, North Ryde, NSW 1670, Australia}
\altaffiltext{6}{School of Physics and Astronomy, University of Southampton, Southampton SO17 1BJ, UK}
\altaffiltext{7}{Center for Astrophysics and Space Astronomy, University of Colorado, 389 UCB, Boulder, CO 80309-389, USA}
\altaffiltext{8}{LPNHE, CNRS-IN2P3 and University of
Paris VI \& VII, 75005 Paris, France}
%\altaffiltext{}{APC, 11 Place Marcelin Berthelot, 75231 Paris Cedex 05, France}
\altaffiltext{9}{Univ. Paris-Sud, Orsay, F-91405, France}
\altaffiltext{10}{Department of Astronomy and Astrophysics, University of
Toronto, 50 St. George Street, Toronto, ON M5S 3H8, Canada}
\altaffiltext{11}{CPPM, CNRS-IN2P3 and University Aix Marseille II, Case 907, 13288 Marseille Cedex 9, France}
\altaffiltext{12}{DSM/IRFU/SPP, CEA-Saclay, F-91191 Gif-sur-Yvette, France}
\altaffiltext{13}{DRDC Ottawa, 3701 Carling Avenue, Ottawa, ON, K1A 0Z4, Canada}
\altaffiltext{14}{Department of Physics and Astronomy, University of
Victoria, PO Box 3055, Victoria, BC V8W 3P6, Canada}


%\altaffiltext{1}{Visiting Scientist, Oskar Klein Centre, University of Stockholm}


\begin{abstract} {We present spectra and lightcurves of \eu\ and \sbv
    , two hydrogen-free superluminous supernovae discovered by the
    Supernova Legacy Survey.  At $z=1.588$, \eu\ is the highest
    redshift superluminous SN with a spectrum, at $M_U=-22.7$ is one
    of the most luminous SNe ever observed, and gives a rare glimpse
    into the restframe ultraviolet where these supernovae put out
    their peak energy.  \sbv\ does not have a host galaxy redshift,
    but based on the supernova spectrum, we estimate it to be at
    $z\sim 1.5$.  Both supernovae have similar observer-frame $griz$
    lightcurves, which map to restframe lightcurves in the $U$-band
    and UV, rising in $\sim 20$ restframe days or longer, and
    declining over a similar timescale.  The lightcurves peak in the
    shortest wavelengths first, consistent with an expanding blackbody
    starting near 15,000 K and steadily declining in temperature.  We
    compare the spectra to theoretical models, and identify lines of
    \ion{C}{2}, \ion{C}{3}, \ion{Fe}{3}, and \ion{Mg}{2} in the
    spectrum of \eu\ and SCP 06F6, and find that they are consistent
    with an expanding explosion of only a few solar masses of carbon,
    oxygen, and other trace metals.  Thus the progenitors appear to be
    related to those suspected for SNe Ic.  A high kinetic energy,
    $10^{52}$ ergs, is also favored.  Normal mechanisms of powering
    core-collapse or thermonuclear supernovae do not seem to work for
    these supernovae.  We consider models powered by \Ni\ decay and
    interaction with circumstellar material, but find that the
    creation and spin-down of a magnetar with a period of 2ms,
    magnetic field of $2 \times 10^{14}$ Gauss, and a 3 \Msun\
    progenitor provides the best fit to the data.}
\end{abstract}





\section{Introduction}
\citet{2009ApJ...690.1358B} announced the discovery of an unusual,
unexplained transient in the course of the Supernova Cosmology Project
cluster supernova search.  With no apparent host galaxy, the redshift
was undetermined.  The event had spectra with broad, unexplained
absorption lines, and a lightcurve in the observed $i$ and $z$ bands
that rose to maximum over 100 days and declined on a similar
timescale.  Soon after, we noticed two similar, puzzling transients in
the Supernova Legacy Survey \citep[SNLS;][]{2010AJ....140..518P},
  \eu\footnote{Each
  SNLS candidate is given a name that consists of the name of the
  survey, the year it was discovered, the field in which it was found
  and a running index consisting of two letters. Hence \eu, was the
  125th candidate to be found in the D4 field during
  2006.} and \sbv , with rise times in $i$ and $z$ bands $> 80$ days, no
obvious host or redshift, and unexplained, broad-lined spectra.
%\footnote{See talk at
%  Stellar Death and Supernovae, 2009, KITP:
% http://online.kitp.ucsb.edu/online/sdeath_c09/howell/}

The key to decoding these mystery objects was found with the discovery
of several new unexplained transients in the course of the Palomar
Transient Factory \citep[PTF; ][]{2009PASP..121.1395L}.
\citet{2011Natur.474..487Q} determined that the new PTF transients and
SCP 06F6 were related to SN 2005ap, an exceptionally luminous
supernova previously discovered during the course of the Texas
Supernova Search at $z=0.2832$ that reached an absolute magnitude
(unfiltered) of -22.7 \citep{2007ApJ...668L..99Q}.  By coadding all
the spectra of SCP~06F6, \citet{2011Natur.474..487Q} detected weak
host galaxy \ion{Mg}{2} lines, and determined the redshift to be
$z=1.189$.  Thus the mysterious lines in the optical spectra of
SCP~06F6 had never before been seen because they were from the
restframe ultraviolet of the supernova.  Furthermore, this explained
the puzzling, seemingly physically implausible long rise
\citep{2009ApJ...704.1251C} of the lightcurve of SCP~06F6 --- it was
due to $1+z$ time dilation.

Key to the link between SN~2005ap and SCP~06F6 were four similar
transients discovered by PTF: PTF09cnd, PTF09cwl, PTF09atu, and
PTF10cwr \citep{2011Natur.474..487Q}.  Because some were at
intermediate redshifts, they had spectra bridging the restframe
optically-dominated SN~2005ap and the restframe UV-dominated SCP~06F6.
They also had a rise to maximum light of 20-50 days, had no hydrogen
in the spectra, were UV-bright, and were exceptionally luminous, with
peak restframe $u$-band (AB) magnitudes $< -21$.  One of these
supernovae, PTF10cwr, also known as SN~2010gx, was also followed by
the Pan-STARRS1 (PS1) survey \citep{2010ApJ...724L..16P}.  The
time-series spectra from \citet{2011Natur.474..487Q} and
\citet{2010ApJ...724L..16P} reveal that, though early-time spectra of
this class of objects do not resemble hydrogen and helium-poor SNe Ic,
late-time spectra do.  At early times, spectra reveal a blue continuum
with a quintuplet of \ion{O}{2} lines in the restframe optical, and
lines tentatively identified as \ion{Si}{3}, \ion{Mg}{2}, and
\ion{C}{2} in the UV \citep{2011Natur.474..487Q}.  Spectra and
lightcurves are consistent with a blackbody with $T \sim 15000$ K at
early times, cooling as it expands.

The extraordinary luminosity of these events, of order 10 times brighter than a
thermonuclear SN Ia, and 100 times brighter than a typical
core-collapse supernova, is difficult to explain.  The lightcurve of a
SN Ia is powered by the radioactive decay of $\sim 0.6$ \Msun\ of \Ni\ 
synthesized in the explosion.  Core-collapse supernovae are generally
powered by a combination of smaller amounts of radioactive
material and the energy of gravitational collapse deposited in what
were formerly the outer layers of the expanding star.  Neither of
these mechanisms seems capable of explaining the extreme luminosity of
these supernovae.  In the first case, their lightcurves do
not have the characteristic rise and fall time associated with \Ni\
decay.  In the second, since the energy is deposited in a small
radius, which is then lost to adiabatic expansion, such extreme
luminosities at relatively late times are not possible.

One class of core-collapse supernovae, SNe~IIn, can attain exceptional
luminosities ($M_V < -21$) by the interaction between the ejecta and
pre-existing circumstellar material cast off by the progenitor.  This
is possible because there is additional energy input at relatively
late times, after the ejecta has expanded.  However, the telltale
signature, narrow hydrogen lines in the spectra, are not seen for the
class of superluminous supernovae covered here.  One possibility is that
the previous cast-off material is not hydrogen-rich, but is instead a
shell of carbon and oxygen lost in a previous pulse of a pulsational
pair-instability supernova \citep{2007Natur.450..390W,
  2011Natur.474..487Q}, although the existence of such phenomena is
speculative.

Another possibility for powering these SNe comes from the
birth and spin-down of a magnetar \citep{2010ApJ...717..245K,
  2010ApJ...719L.204W}.  In this model, a highly magnetized, rapidly
spinning neutron star is created in the supernova.  It rapidly loses
energy, decreasing its spin period, and transfers some of it to the
expanding supernova ejecta. While the required energies are achievable,
it is not clear how the energy from the spin-down would be coupled to
the nascent supernova, nor (until now) what the spectra or lightcurves of such
events would look like.  

We can hope to gain some insight into the physical mechanism
surrounding superluminous supernovae by examining their stellar
environments.  \citet{2011ApJ...727...15N} found that these events
generally favor UV-bright, blue hosts with a low mass and high star
formation rate.  For a review of all properties of superluminous SNe
(SLSNe), including hydrogen-rich events, see
\citet{2012Sci...337..927G}.

Here we present data on the two superluminous SNe with spectra from
the SNLS, SNLS~06D4eu and SNLS~07D2bv. One of these,
06D4eu, has an accurate redshift from its host galaxy, allowing an
accurate determination of its luminosity.  Therefore, we use it as 
a case study, comparing it to theoretical models, including new work
on the magnetar spin-down model of \citet{2010ApJ...717..245K}.
The paper is divided as follows. In section 2, we summarize key
aspects of the SNLS survey and describe how two superluminous SNe,
\eu\ and \sbv , were discovered there.  In section 3, we present
VLT/XSHOOTER observations that enable us to derive the redshift and
star formation rate of the likely host of \eu.  In section 4, we
present the lightcurves and fit them with blackbody models.  In
section 5 we present the spectra and fit them with a radiative transfer
model to identify features.  In section 6 we try to explain the
luminosity of \eu\ using models, and we discuss progenitor scenarios.
Finally, we conclude in section 7.  Throughout the paper we use Vega
magnitudes unless otherwise noted and we assume a flat
lambda-dominated Universe with $\Omega_{\mathrm{M}}=0.27$ and
$H_{0}=71$ km s$^{-1}$ Mpc$^{-1}$.

\section{Discovery}

\subsection{The Supernova Legacy Survey}

The Supernova Legacy Survey (SNLS) was designed to
constrain the dark-energy equation-of-state parameter using distances and redshifts of several hundred spectroscopically
confirmed Type Ia supernovae (\sneia). The SNLS was comprised of two
parts: an imaging part that was based on the Deep Survey of the
Canada-France-Hawaii Telescope Legacy Survey\footnote{http://www.cfht.hawaii.edu/Science/CFHLS/} (CFHT-LS), and
a spectroscopic part that was designed to follow-up supernova candidates
that were discovered in the imaging survey.

The imaging survey consisted of regular imaging of 4 fields, labeled
D1 to D4, in the four filters: {\it g}, {\it r}, {\it i} and {\it z}.
Each field was observed once every 3 to 5 nights on and around new
moon. This technique for finding transients, which is often referred
to as the ``rolling search'' technique, naturally leads to
well-sampled, multi-colored light-curves.  Over the five years during
which the survey ran, the lightcurves of approximately 1000 \sneia\ and
many other variable sources were obtained. Details of the real-time
observing strategy and follow-up can be found in
\citet{2010AJ....140..518P}.  Three year results are published in
\citet{2010A&A...523A...7G}, \citet{2011ApJS..192....1C}, and
\citet{2011ApJ...737..102S}.

The spectroscopic follow-up for the SNLS was done with several
instruments: GMOS on Gemini North and Gemini South
\citep{2005ApJ...634.1190H,2008A&A...477..717B,2011MNRAS.410.1262W},
LRIS and DEIMOS on the Keck I and Keck II telescopes
\citep{2008ApJ...674...51E}, and FORS1 and FORS2 on VLT Antu and
Kueyen \citep{2009A&A...507...85B}.  

Lightcurves and spectra from this paper will be made available via the
WISeREP repository \citep{2012PASP..124..668Y}.

\subsection{\eu}
\begin{figure}
\plotone{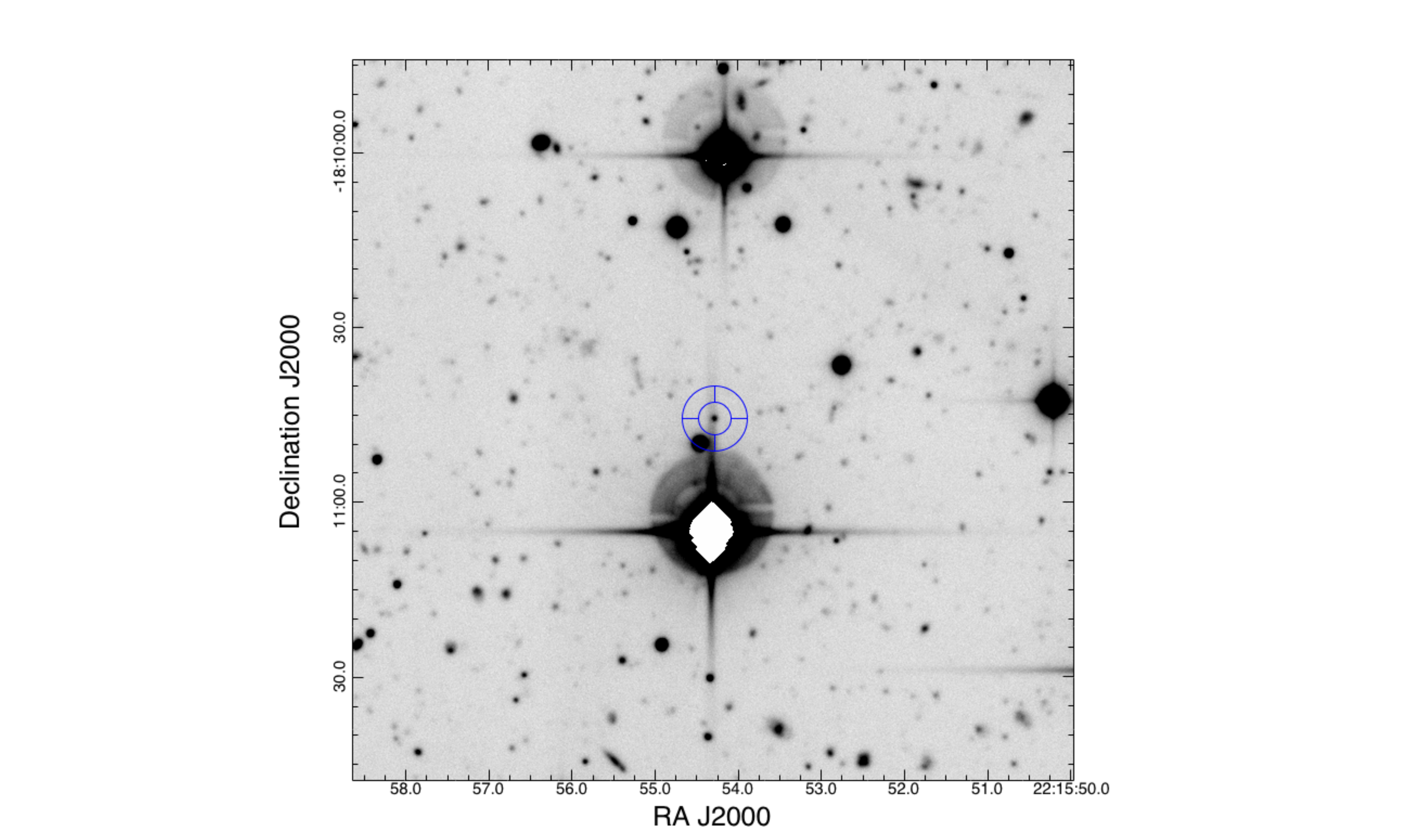}
%\plotone{06D4eu_triplet.pdf}
\plotfiddle{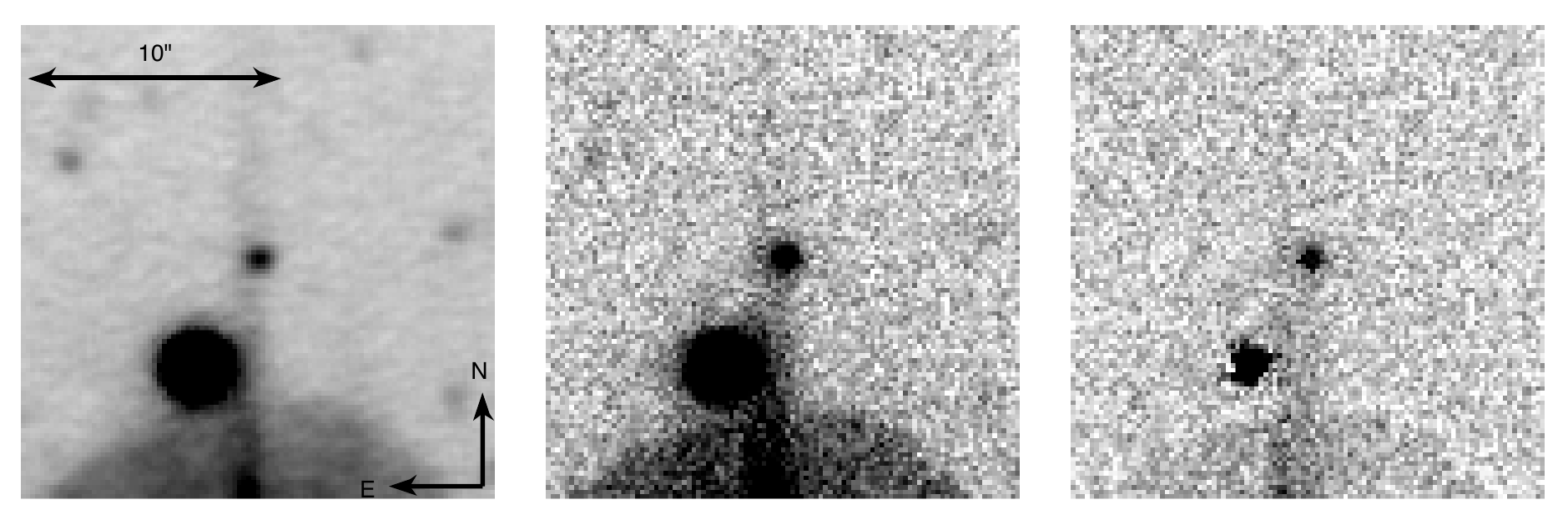}{0in}{0}{300}{100}{0}{0}

\caption{{\it Above:} CFHT MegaCam image centered on \eu. This cut-out is 2 arc minutes on a side and north is up and east
  left. The position angle of the X-shooter slit was aligned so that
  scattered light from the 13th magnitude star that is only
  20\arcsec\ south of \eu\ did not contaminate the spectrum. {\it
    Below}. The image detection triplet, which from left to right shows
  the reference image, an image of
  \eu\ when it was close to maximum light, and the difference between
  the first two. The images were taken in the {\it
    g}-band. \label{fig:image}}
\end{figure}

\eu\ was discovered images taken August 20, 2006 (UT), processed with
the real-time processing pipeline \citep{2010AJ....140..518P}, at
$i=23.7$. It was subsequently recovered on prior imaging going back to
July 27, 2006.  The field of \eu\ is shown in Fig.~\ref{fig:image}.

As part of the regular follow-up of SNLS candidates, \eu\ (RA 22:15:54.291 DEC
-18:10:45.56 J2000) 
was submitted to the ESO VLT for spectroscopic confirmation.  At this
stage, it was flagged as a rank B candidate, i.e., it was thought to be a
\snia\ based on early pre-maximum light-curve data, but there
were some anomalies in the early part of the lightcurve that prevented
it from being classified as a rank A candidate.

\begin{figure}
\plotone{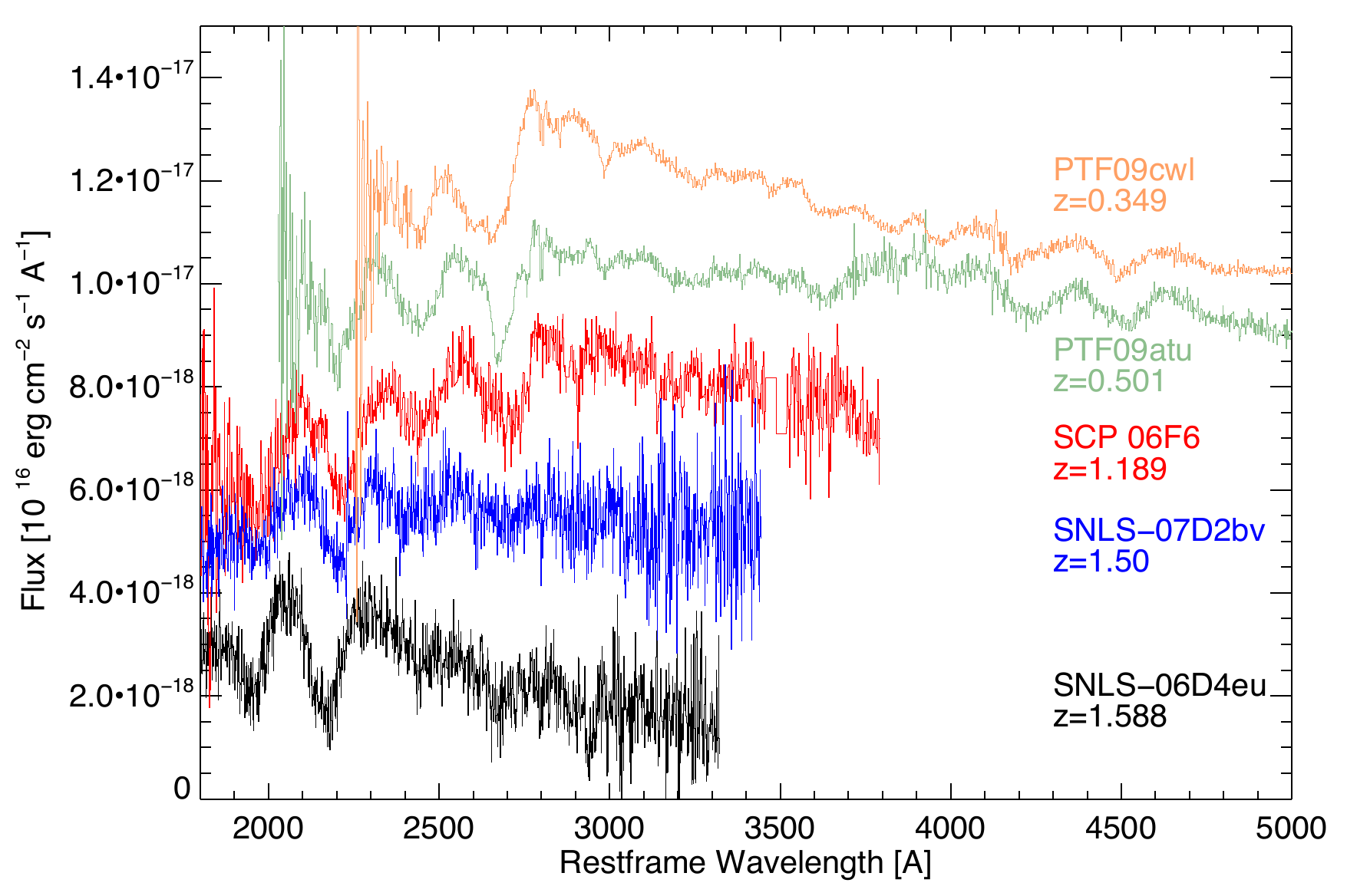}
\caption{Spectra of \eu\ and \sbv\ compared to other similar events in
  the restframe \citep{2011Natur.474..487Q,2009ApJ...690.1358B}.  \eu\ extends farthest to the restframe UV because it
  is as the highest redshift.  The overall shape of the spectrum also
  peaks near 2000 \AA , in contrast to the other events.  This bluer
  color is likely because the spectrum was taken earlier than the
  other events.  Note that the absorption feature near 2200 \AA\ is
  strongest in \eu .  Additionally, an absorption feature near 1900
  \AA , seen in SCP 06F6, is apparent in \eu . \sbv\ is shown at a
  redshift of $z=1.50$, although this is approximately determined from
  matching its features with \eu ; no host galaxy redshift could be
  obtained.  We show the second of three spectra obtained for SCP 06F6.  The first does not extend far enough to the blue to make a meaningful comparison, and the third is not substantially different from the second.\label{fig:spec}}
\end{figure}

Soon after the request was submitted, \eu\ was observed with 
the MOS (Mutli-Object Spectroscopy) mode of FORS1 on August 31, 2006.  
FORS1 was used with the 300V grism in conjunction with the GG435 order
sorting filter.  The slit width was set to 1\arcsec, which results in
a spectral resolution of 400.
%For slits that were located near the
%centre of the 7\arcmin\ $\times$ 7\arcmin\ FORS2 field-of-view, 
The resulting spectrum (Fig.~\ref{fig:spec}) starts around 4250 \AA\ and ends at
approximately 8500 \AA. The exposure time was 4500 seconds and was
split into 5 exposures of 900 seconds each.

The FORS1 data were processed with standard IRAF
routines\footnote{IRAF is distributed by the National Optical
  Astronomy Observatories, which are operated by the Association of
  Universities for Research in Astronomy, Inc., under the cooperative
  agreement with the National Science Foundation.}. The detector bias
was removed by fitting a low order polynomial to the overscan region.
Pixel-to-pixel sensitivity variations were removed using an image of
uniformly-illuminated white screen. The spectrum of \eu\ was then
extracted and calibrated in wavelength and flux.

The advantage of the MOS mode is that, in addition to obtaining a
spectrum of \eu, the MOS mode allows us to target other objects such
as the host galaxies of transients, many of them supernovae, for which
we could not get a spectrum when they were bright. Towards the end of
the SNLS project, typically 5 of the 19 MOS slits could be placed on
such targets. The remaining slits were placed on randomly chosen field
galaxies.

The MOS data were processed one month after the data were taken;
however, we failed to appreciate the interesting nature of the SN spectrum
until 8 months later (no such supernova had ever been identified at
that point), so we lost an opportunity to follow it spectroscopically
and at other wavelengths.

\begin{figure}
\plotone{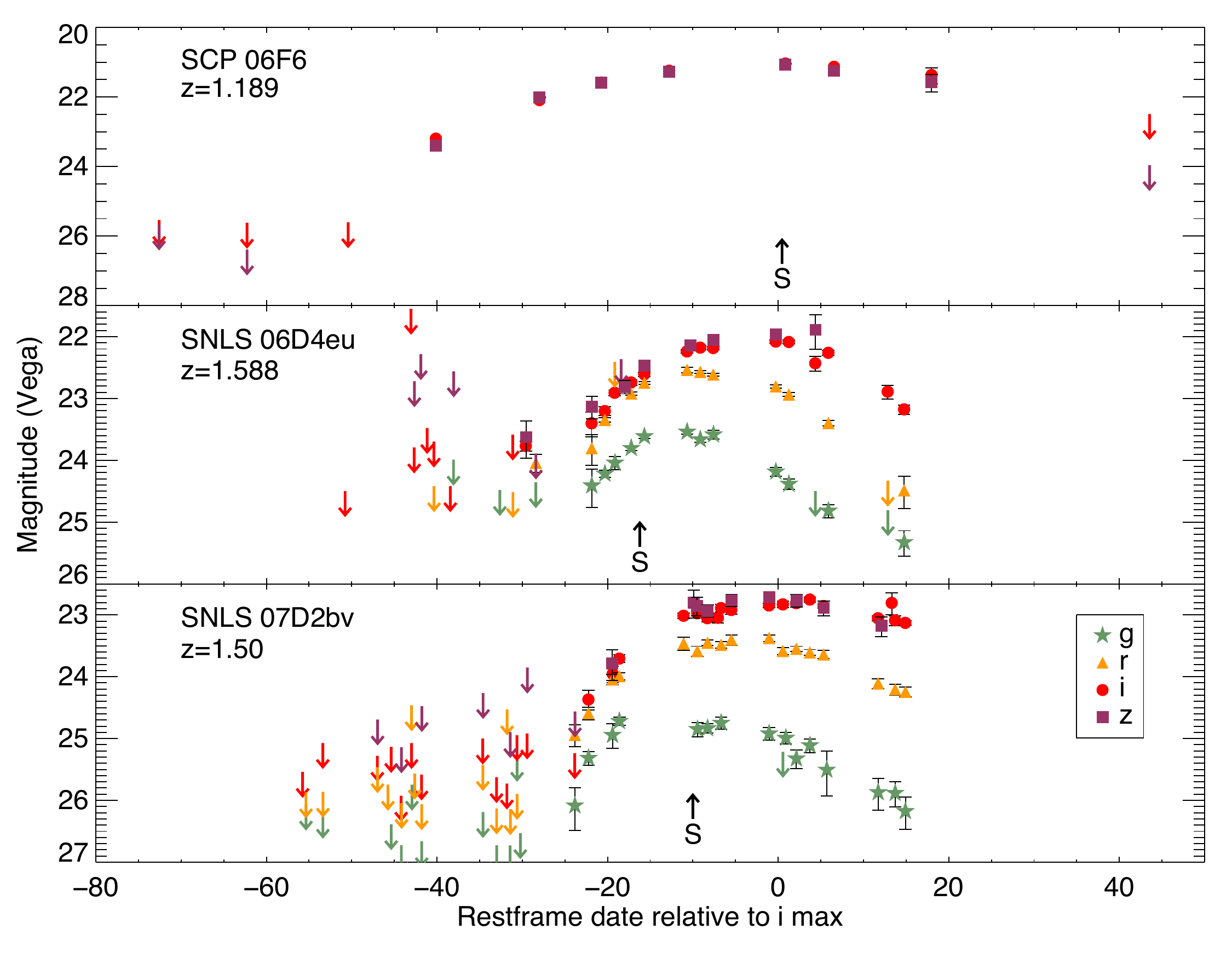}
\caption{Lightcurves of SCP 06F6, \eu , and SNLS~07D2bv.  Observed  
magnitudes are shown in the Vega system, although the time axis has 
been corrected for $1+z$ time dilation.  An `S' with an arrow denotes 
when the spectra shown in Fig.~\ref{fig:spec} were taken. Downward arrows denote 3$\sigma$ upper limits.
\label{fig:phot}}
\end{figure}

Fortunately, SNLS was a rolling search, meaning that each field was
repeatedly imaged to both search for new supernovae while building
lightcurves of previously detected events.  Therefore, as for all
transients, $griz$ lightcurves of \eu\ were obtained by default, and
could be analyzed in retrospect.  This lightcurve is shown in
Fig.~\ref{fig:phot}, and discussed in detail in
Section~\ref{lightcurves}.

Just as for SCP 06F6, from the initial spectrum (Fig.~\ref{fig:spec}), the redshift of \eu\
was unclear.  We attempted to spectroscopically observe the host
galaxy in the optical after the SN faded, but did not detect host
galaxy lines.  Ultimately, a multi-hour VLT X-shooter spectrum showed
emission lines in the IR, revealing the redshift to be 1.588 (Section
\ref{euhost}).  

\subsection{SNLS~07D2bv}
\begin{figure}
\plotone{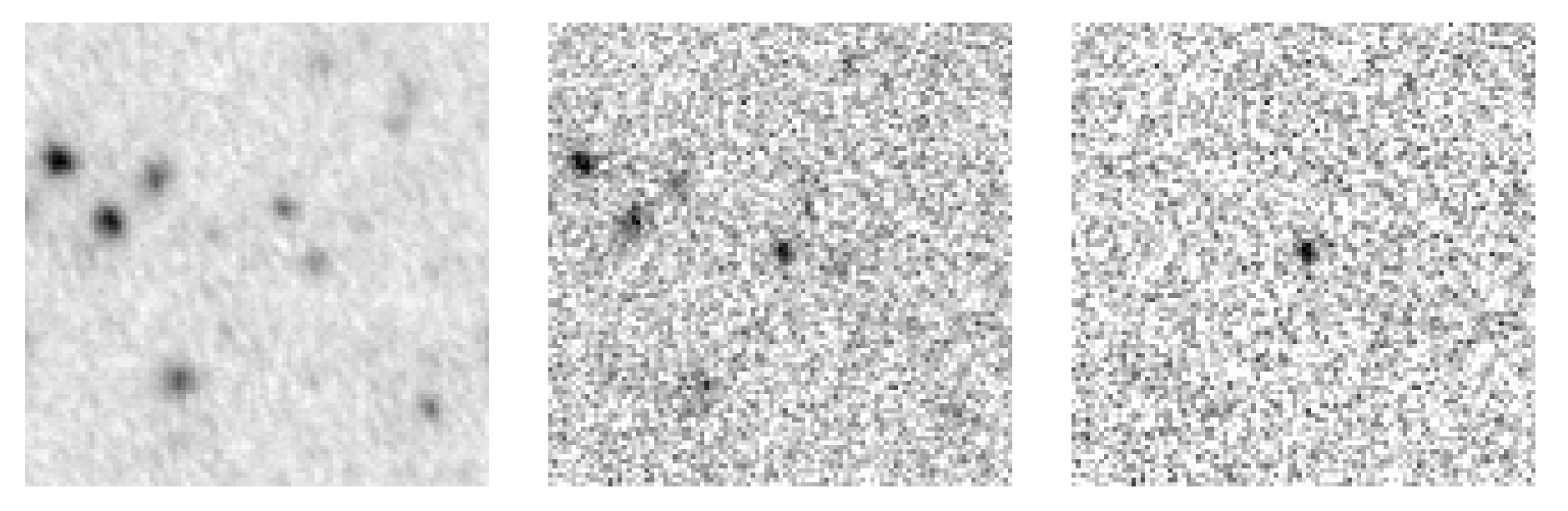}
\caption{Detection triplet for SNLS~07D2bv.  The leftmost image is the reference image, created by stacking SNLS data from years in which the supernova did not appear.  The middle image shows the supernova (center) after it started to brighten.  The right image is a subtraction showing the discovery of the supernova.  The scale is the same as in Fig.~\ref{fig:image}
\label{fig:07D2bvtriplet}}
\end{figure}

SNLS~07D2bv (Fig.~\ref{fig:07D2bvtriplet}) was discovered February 22, 2007 by the SNLS (at $i=24$ according to real-time photometry),
but was then detected at $5\sigma$ or greater on prior images
beginning on Feb.  9.  The lightcurve is shown in Figure~\ref{fig:phot}.

It was categorized as a class `C' candidate, meaning it was unlikely
to be a SN Ia, but could be one.  As a result of its low priority, a
spectrum was not obtained until March 16, when it was observed by VLT
FORS1 in MOS mode, by which time the SN had risen to $\approx 23$ mag
in $z$-band.  The spectrum is shown in Figure~\ref{fig:spec}.  The
setup and data reduction are identical to that mentioned above for \eu
, but the exposure time was $4 \times 900$s.  The one-hour spectrum
revealed broad SN-like features, with no apparent host lines, and was
eventually noted as ``06D4eu-like.''  Due to the faintness of the four
nearby potential host galaxies, it was not possible to obtain
subsequent host galaxy spectroscopy.

\section{The host of \eu \label{euhost}}
\subsection{Observations with X-Shooter}

Spectra of the host of \eu\ covering 300--2500\,nm 
were obtained with X-Shooter/Kueyen during 2009 and 2010.  X-Shooter
\citep{2006MNRAS.372.1333D} consists of three arms: an
ultraviolet-blue arm (UVB) covering the 300--550\,nm spectral region,
a visible arm (VIS) covering the 550--1000\,nm spectral region and a
near-infrared arm (NIR) covering the 1000--2500\,nm spectral
region. Each arm has its own cross-dispersed spectrograph and
detector. The light from the target is split between the three arms by
two dichroics. The configuration of X-Shooter that was used to observe
the host of \eu\ is listed in Table \ref{tab:Xobs}.

The position angle of the instrument was set so that scattered light
from the bright star that is located 20\arcsec\ to the South of
\eu\ (see Fig.~\ref{fig:image}) would not contaminate the spectrum of
the host of \eu.  During the observations, the sky was clear and the seeing, as
measured by the site seeing monitor, averaged around 0\farcs8.

\begin{table}
\begin{center}
\caption{X-Shooter set up\label{tab:Xobs}}
\begin{tabular}{lcccc}
  \tableline\tableline
  Arm & Slit  & Res. \tablenotemark{a} & Integration & Integration\\
      & width $\times$ length   &      &  (2009) [s]   & (2010) [s] \\
  \tableline
  UVB &  1\farcs0 $\times$ 11\arcsec & 5000 & 9000 & 7200\\
  VIS &  0\farcs9 $\times$ 11\arcsec &  9000 & 9000 & 7200\\
  NIR &  0\farcs9 $\times$ 11\arcsec &  5000 & 8640 & 7200\\
  \tableline
\end{tabular}
\tablenotetext{a}{The resolution is set by the slit.}
%\tablenotetext{b}{One of the exposures in the NIR arm failed, so the
%  observing sequence was partially repeated, thus resulting in seven
%  480 second exposures in the NIR arm.}
\end{center}
\end{table}

To facilitate the removal of the NIR background and to mitigate the
effect of the large number of hot pixels in the NIR detector, the
target was offset between two positions along the echelle slit. During
2009, the frequency of the offsetting was once every $\sim1600$
seconds, during which three 480 second exposures in the NIR arm and
one 1500 second exposure in each of the UVB and VIS arms were
taken. Already, from the data taken in 2009, lines from [OII], [OIII]
and H$\alpha$ could be detected in the data.  However, H$\beta$ landed
in a region that was contaminated with bright night sky lines, so a
secure detection of H$\beta$ could not be made. At that stage, the
target was only visible during the beginning of the night, which is
when night sky lines are bright and variable. We suspended the program
for six months, so that the target could be observed towards the end
of the night when night sky lines are considerably fainter. We also
increased the offset frequency to once every $\sim 300$ seconds and
took one 300 second exposure in each of the arms at each offset
position. These data were taken in 2010. The increased offset
frequency allowed us to remove the OH and O$_{2}$ lines near H$\beta$
better. The visibility of H$\beta$ was also helped by Earth's motion
around the sun. The resulting Doppler shift moved H$\beta$ away from
the brightest O$_{2}$ line. The total integration times for the 2009
and 2010 observations are reported in Table~\ref{tab:Xobs}.

Both the VIS and NIR arms were processed in an identical
manner. Consecutive images were grouped into pairs, which consisted of
images that were taken on either side of the offset. Within each pair,
one image was subtracted from the other. This results in a frame in
which the object appears twice: once with positive counts and again
with negative counts. While this method removes most of the
background, significant residuals remain, and an additional step to
remove these residuals was necessary.

The tilt of the path that traces the spatial direction changes
noticeably as one moves along an order, and this adds some complexity
to the removal of the residuals. At this point, one can choose to
rectify the spectrum and proceed with removing the residuals or, as we
do, compute the residual along a path that follows the spatial
direction. For this, we used our own custom made software.  The latter
avoids interpolation, which introduces correlations between pixels and
tends to grow the region over which bad pixels, which are quite common
with the current NIR detector of X-Shooter, can affect the data. For
the UVB arm, the sky background was removed from all images directly.

Apart from the removal of the sky background, the processing of the
data was standard and we used IRAF for most of the processing
tasks. For all arms, two spectra, offset by the nod throw, were
extracted. All spectra were then wavelength calibrated.  Where there
were sufficient night sky lines (all of the NIR and most of the VIS),
we used night sky lines to do the wavelength calibration.  Otherwise
we used arc frames. A star with spectral type B9V was used to remove
telluric features and to calibrate the flux scale. As a final step,
all spectra were then corrected to the heliocentric reference frame.

\subsection{The redshift of the host of  \eu}

The X-shooter spectrum of the host \eu\ shows several clear emission lines, such
as [\ion{O}{2}] $\lambda\lambda$\,3727,3729, H$\beta$ [\ion{O}{3}]
$\lambda\lambda$\,4959,5007 and H$\alpha$. They lead to a host
redshift of $z=1.5881 \pm 0.0001$\footnote{The redshift is in the
  heliocentric frame.}.  The [\ion{O}{2}] doublet is resolved and both
components of the [\ion{O}{3}] doublet are detected. The fluxes of all
lines are listed in Table~\ref{tab:linefluxes}.  We tentatively detect
[\ion{N}{2}] $\lambda$\,6584 in the 2-d sky-subtracted data.  However,
the detection is marginal (only 4--$\sigma$) and the relative 
uncertainty in the derived flux is large. 

\begin{figure}
\plotone{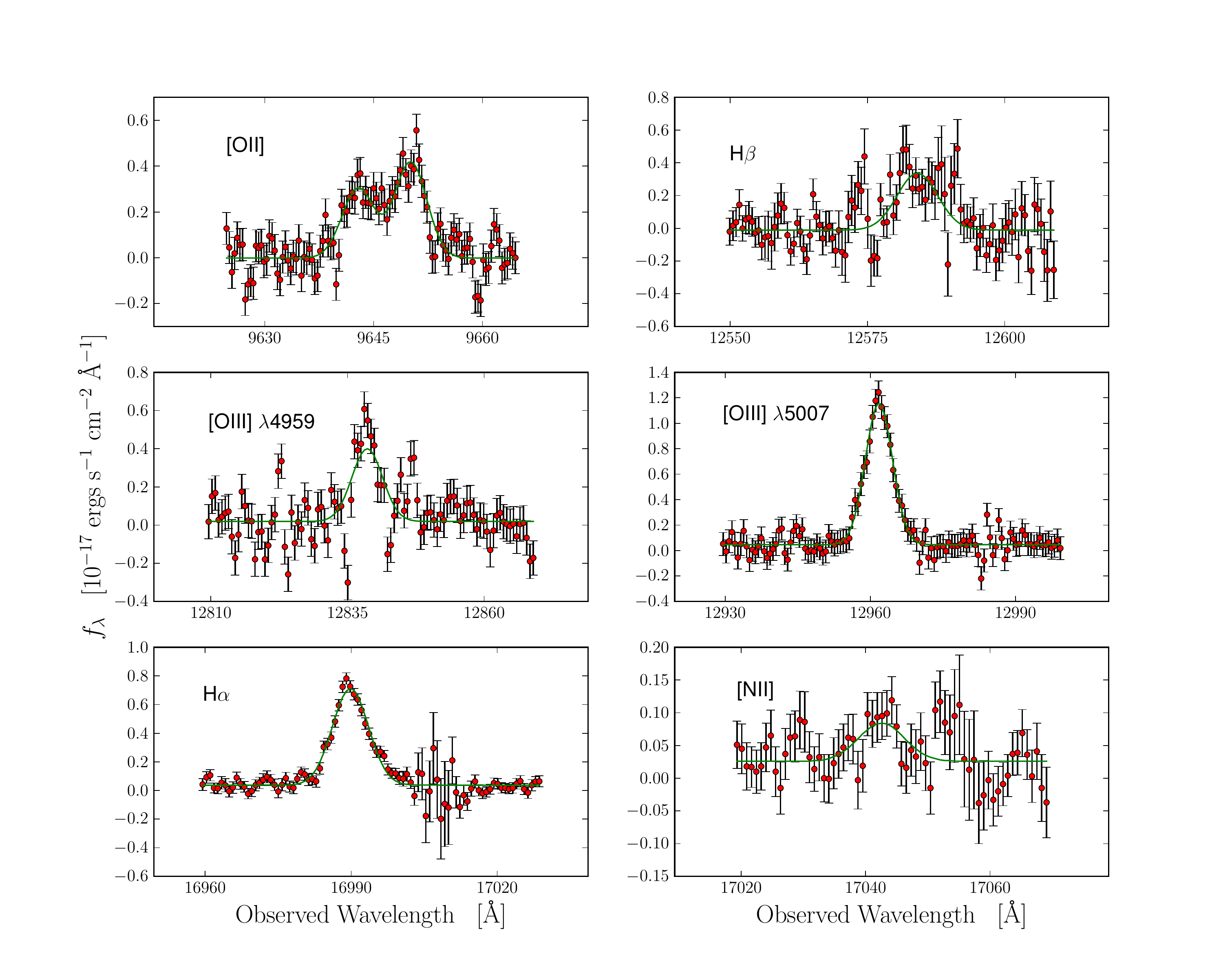}
\caption{The six emission lines detected in the host galaxy of
  SNLS 06D4eu together with a Gaussian fit (green line). The [OII]
  $\lambda\lambda$ 3727,3729 doublet is fitted with two Gaussians. The
  line fluxes reported in Table~\ref{tab:linefluxes} are determined by
  integrating the the area under the fits.  \label{fig:lines} }
\end{figure}

%The [\ion{S}{2}] doublet is not detected.
%Most of these undetected lines land close to night sky
%lines, so the upper limits to the line fluxes are difficult to
%estimate. The upper limits for H$\beta$ and [\ion{N}{2}]
%$\lambda$\,6584 shown in Table~\ref{tab:linefluxes} should be used
%with this caveat in mind.

Most lines land in regions that are relatively free of bright night
sky lines and telluric features. The [\ion{O}{2}]
$\lambda\lambda$\,3727,3729 doublet is slightly contaminated by both,
but neither the sky lines nor the telluric absorption features near
the doublet are strong enough to bias the flux measured for
[\ion{O}{2}]. H$\beta$ lands close to a bright night sky line.  It is
more clearly separated from the line in the 2010 data, thanks to the
motion of the Earth around the Sun, so we only use the 2010 data when
estimating the H$\beta$ line flux. The continuum is detected, but only
in the VIS and UVB arms. It is clearest in the UVB arm; however, the
signal-to-noise ratio is low and absorption features cannot be seen.

The lines detected in the X-shooter spectrum land in regions that lie
outside the wavelength intervals that were covered in earlier attempts
to get the host redshift using GMOS on Gemini South and FORS2 on the
VLT.

\begin{table}
\begin{center}
\caption{Observed line fluxes of the host of \eu. \label{tab:linefluxes}}
\begin{tabular}{ll}
\tableline\tableline
Line                              	   & Observed line flux \\
                                  	   & ($10^{-17}\rm{erg/s/cm}^2$)\\
\tableline
\,[\ion{O}{2}] $\lambda\lambda$\,3727,3729 & $4.0^{+0.6}_{-0.7}$  \\
\,H$\beta$                                 & $3.3^{+0.4}_{-0.3}$  \\
\,[\ion{O}{3}] $\lambda$\,4959 		   & $7.8^{+0.7}_{-0.5}$  \\
\,[\ion{O}{3}] $\lambda$\,5007 		   & $2.7^{+0.3}_{-0.3}$  \\
\,H$\alpha$                   		   & $6.3^{+0.4}_{-0.4}$  \\
\,[\ion{N}{2}] $\lambda$\,6584 		   & $0.55^{+0.10}_{-0.15}$  \\
\tableline
\end{tabular}

\end{center}
\end{table}

{
\begin{table}
\begin{center}
\caption{Derived quantities from host lines. \label{tab:metal}}
\begin{tabular}{ll}
\tableline\tableline
Indicator                              	   & Value \\
\tableline
\,H$\alpha$/H$\beta$                        & $1.9^{+0.3}_{-0.2}$  \\
\,SFR(H$\alpha$)                           & $8.3\pm0.3$ \Msun\ yr$^{-1}$ \\
\,[\ion{N}{2}]/[\ion{O}{2}]                & $0.14\pm 0.04$  \\
\,[\ion{N}{2}]/H$\alpha$                   & $0.09\pm 0.02$  \\
\,$R_{23}$                 		   & $4.4^{+0.6}_{-0.5}$  \\
\,$O_{32}$                 		   & $2.6^{+0.6}_{-0.4}$  \\
\,12+log(O/H) [lower]                      & 7.9  \\
\,12+log(O/H) [upper]                      & 8.9  \\
\,[\ion{N}{2}] $\lambda$\,6584 		   & $0.55^{+0.10}_{-0.15}$  \\
\tableline
\end{tabular}

\end{center}
\end{table}
}

%With an accurate measure of the host redshift, we searched for
%absorption lines in the FORS spectrum of \eu\ coming from the
%interstellar medium (ISM) of the host. We detect what may be a very
%weak MgII feature at z=1.586; however, the signal-to-noise ratio is
%very low, so the identification of this feature with MgII is not
%certain. If real, the slight blueshift of the line with respect to
%lines coming from the ionised gas is indicative of an outflowing
%galactic wind with a velocity of $\sim 200\,\mathrm{km\,s}^{-1}$.

% Quote an EW
% Show it as an inset

\subsection{Dust extinction, star formation rate and metallicity}
 Derived quantities from host galaxy lines are summarized in Table~\ref{tab:metal}.

An estimate of the amount of extinction from dust can be made by
comparing the measured H$\alpha$ to H$\beta$ line ratio with the
theoretical value, which for case-B recombination is 2.86
\citep{1989agna.book.....O}. We measure $1.9^{+0.3}_{-0.2}$, a value
which is slightly lower than the theoretical value. The difference is
opposite to what one expects to occur if there is extinction from
dust, so we conclude that there is no evidence for significant
extinction and we posit that the emission lines and the quantities
computed from them are unaffected by dust.

Following \citet{1998ARA&A..36..189K}, we use the H$\alpha$ line flux to
derive the star formation rate (SFR) of the host of \eu.  The
H$\alpha$ flux translates to a SFR of  $8.3\pm0.3$ \Msun\ yr$^{-1}$, which
means that the host of \eu\ is vigorously forming stars. 

Estimating the metallicity of the gas from emission lines is more
problematic.  The \rindex\ index \citep{1979MNRAS.189...95P}, which uses [\ion{O}{2}],
[\ion{O}{3}] and H-beta, is sensitive to the gas phase metallicity and
$q$, the ionization parameter \citep{2004ApJ...617..240K}.  The index
is also doubly valued, i.e., for one value of \rindex, two
metallicities are possible. The two values are usually referred to as
the upper and lower branches. Hence, using it to derive the
metallicity requires one to decide which branch to use. This requires
additional information, such as the [\ion{N}{2}]\,$\lambda$\,6584 to
H$\alpha$ line ratio \citep{2004ApJ...617..240K} or the
[\ion{N}{2}]\,$\lambda$\,6584 to [\ion{O}{2}] line ratio
\citep{2008ApJ...681.1183K}, for which we derive $0.09\pm0.02$ and
$0.14\pm0.04$, respectively. The [\ion{N}{2}]/[\ion{O}{2}] ratio
indicates that the upper branch should be used; however, given the
uncertainty in the ratio, we cannot exclude the possibility that the
lower branch should be used instead.  Deeper spectra are needed to
answer this question.

\citet{2004ApJ...617..240K} recommend an iterative approach to computing the
metallicity using both \rindex\ and \oindex. The \oindex\ index is
sensitive to both metallicity and to the ionization parameter. For
the host galaxy of \eu , we find \rindex$=4.4^{+0.6}_{-0.5}$ and
\oindex$=2.6^{+0.6}_{-0.4}$, which corresponds to a gas phase metallicity of
$12+\mathrm{log(O/H)}=8.9 (7.9)$ if the galaxy lands on the upper
(lower) branch\footnote{For reference, the oxygen abundance,
  $12+\mathrm{ log(O/H)}$, is related to the metal mass fraction by
  $Z=29 \times 10^{[12+\mathrm{log(O/H)}]-12}$
  \citep{2004ApJ...617..240K}.}. For reference, the solar value is
$12+\mathrm{log(O/H)} \sim 8.7$.  

We do not attempt to derive a host galaxy mass for \eu , due to a lack
of long wavelength data.  We do not have any data on the host redward
of the 4000 \AA\ break, since all observed optical data corresponds to
the restframe UV.  Furthermore, no observed IR data exists.  The
WIRCAM data of the D4 field \citep{2010A&A...523A..66B} does not cover
the supernova position, and the VISTA hemisphere survey
\citep{2012sngi.confE..37M} has not yet targeted this area.

\section{Lightcurves\label{lightcurves}}

\begin{figure}
\plotone{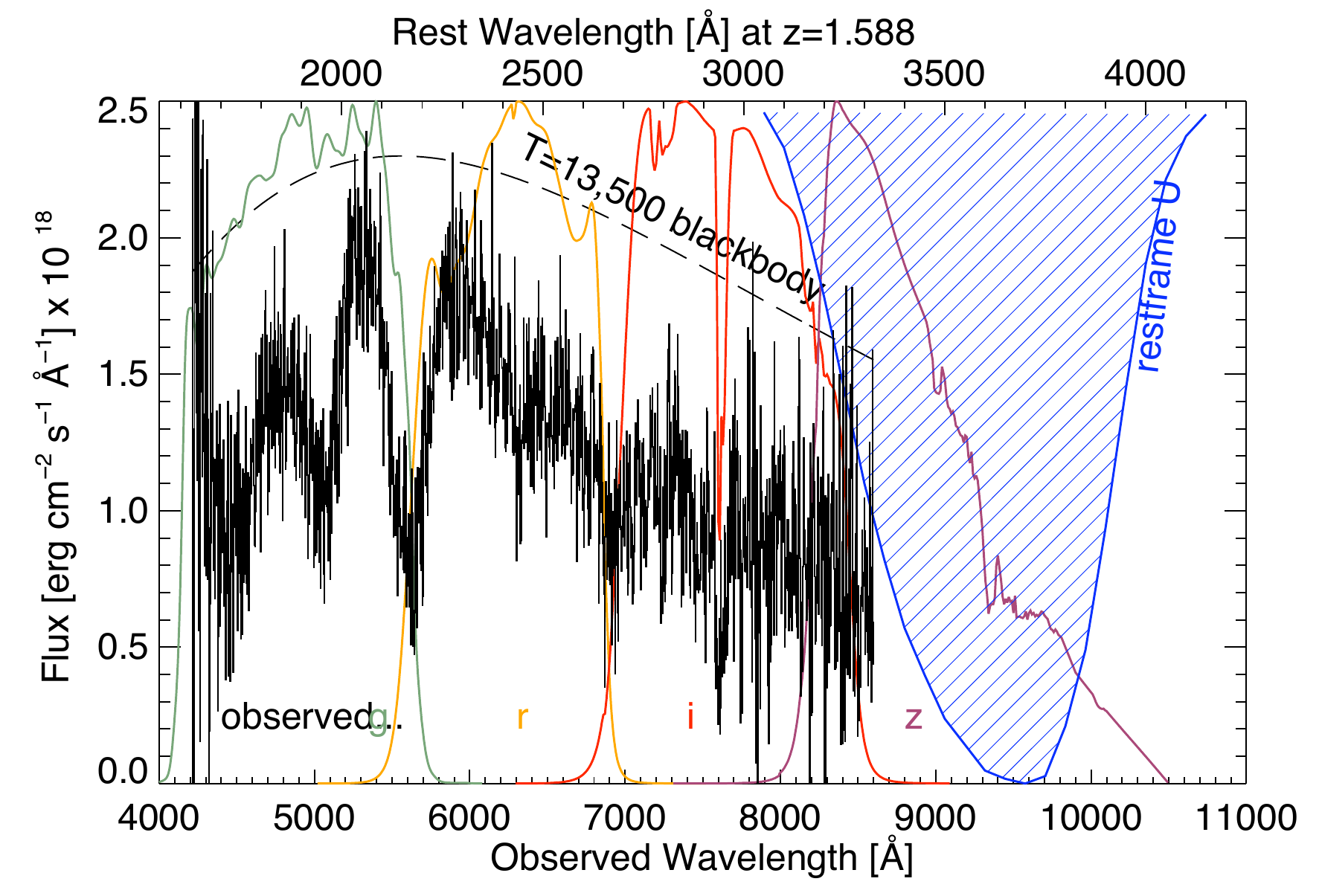}
\caption{Spectrum of \eu\ at observed and restframe
  wavelengths.  CFHT Megacam filter bandpasses, used to collect the
  photometry in Fig.~\ref{fig:phot} are indicated.  Restframe $U$-band is
  plotted upside down to indicate that it corresponds to the upper
  axis.  It is a good match to observed \zp .  Note that \gp , \rp ,
  and \ip\ correspond to restframe ultraviolet regions of the
  spectrum.  The overall shape of the spectrum is consistent with a
  $T=13500$ K blackbody.  This is consistent with and independent
  evidence of the high temperatures indicated in the simple blackbody
  model fit to the photometry as discussed in section
  \ref{blackbody}.\label{fig:bandpass}}
\end{figure}

\begin{figure}
\plotone{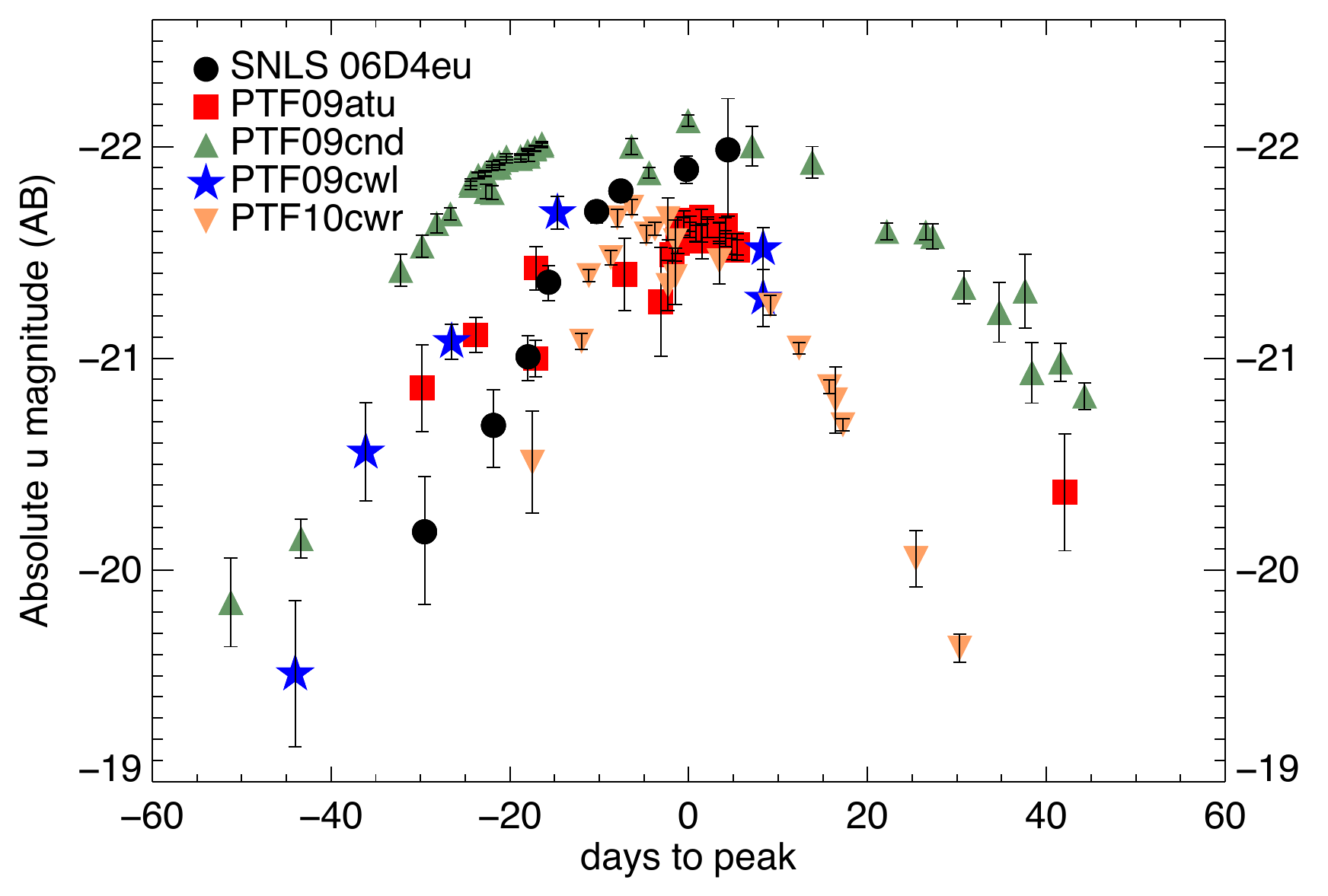}
\caption{ Absolute restframe $u\arcmin$ magnitudes of SLSNe
  in the AB system.  \eu\ data is k-corrected from observed
  Megacam $z$-band data to SDSS $u\arcmin$ using the blackbody 
  fit as the SED.  Comparison SNe are from
  \citet{2011Natur.474..487Q}.  Magnitudes are $\sim 0.7$ brighter
  in the Vega system.  \eu\ is similar to the brightest of the
  SLSNe, but has a shorter rise to maximum light, around 30 days.  
The time of peak is uncertain to about 5 days, due to the large error on the final point.
  The PTF SNe have an uncertainty of several tenths of a magnitude 
  due to k-correction uncertainties (Quimby, private communication).
  \label{fig:uband}
}
\end{figure}

%\begin{table}
%\begin{center}
%\caption{Superluminous SNe\label{tab:SLSNe}}
%\begin{tabular}{ccccccc}
%  \tableline\tableline
%  Supernova & Redshift  & Abs. peak. \tablenotemark{a} & Rad. Energy & Spec. phase & Wavelength & Ref.\\
%      &  &  & [mag]  & erg & days & \AA & \\
%  \tableline
%  SN 2005ap & 0.2832 & -22.73  & $1.2\times 10^{51}$ & & & Q07, Q11\\
%  SCP 06F6  &  & -22.73  & $1.2\times 10^{51}$ & & & \\
%  PS1-10ky  &  & -22.73  & $1.2\times 10^{51}$ & & & \\
%  PS1-10awh &  & -22.73  & $1.2\times 10^{51}$ & & & \\
%  PTF09atu  &  & -22.73  & $1.2\times 10^{51}$ & & & \\
%  PTF09cnd  &  & -22.73  & $1.2\times 10^{51}$ & & & \\
%  SN 2009jh &  & -22.73  & $1.2\times 10^{51}$ & & & \\
%  SN 2006oz &  & -22.73  & $1.2\times 10^{51}$ & & & \\
%  SN 2010gx &  & -22.73  & $1.2\times 10^{51}$ & & & \\
%   &  & -22.73  & $1.2\times 10^{51}$ & & & \\
%   &  & -22.73  & $1.2\times 10^{51}$ & & & \\
%   &  & -22.73  & $1.2\times 10^{51}$ & & & \\
%  \tableline
%\end{tabular}
%\tablenotetext{a}{The resolution is set by the slit.}
%%\tablenotetext{b}{One of the exposures in the NIR arm failed, so the
%%  observing sequence was partially repeated, thus resulting in seven
%%  480 second exposures in the NIR arm.}
%\end{center}
%\end{table}

The observed $\sim 80$ day rise in the $i$ and $z$ band for \eu\ and
SNLS~07D2bv were initially puzzling, as SNe powered by radioactive decay have
a much shorter rise (e.g. $\sim 20$ days for SNe Ia), and those
powered by collisions with circumstellar material stay at a fairly
constant luminosity, e.g. SNe IIn.  This mystery was resolved with the
determination of the redshift of \eu\ --- due to ($1+z$) time
dilation, the rest frame rise time is in fact much shorter, around 30
days (Figure~\ref{fig:phot}) in the reddest observed bands.

In fact, because of the high redshift, the Megacam bands are sampling
exclusively UV and $U$-band light.  Figure~\ref{fig:bandpass} shows
that observer-frame $z$ corresponds to rest-frame $U$ for \eu .
The $g$, $r$, and $i$ lightcurves are sampling restframe band centered
near 1900 \AA , 2400 \AA, and 2900 \AA\ respectively.  Thus,
Figure~\ref{fig:phot} is remarkable in that it shows for the first
time the entire near-UV lightcurve evolution for two superluminous
supernovae.  It is apparent that these supernovae are exceptionally
UV-bright and blue at early times. The SN peaks in the observed $g$
band (rest 1900 \AA ), and $r$-band (rest 2400 \AA ) about 5-12 days
before the restframe $U$ band, earlier for \eu , and later for
SNLS~07D2bv.

Before day -15 the colors of the SNe are fairly constant, but they
rapidly become red as maximum light in the restframe $U$-band is
approached.  After the peak, the colors return to a fairly constant
value.  This can be seen in Figure~\ref{fig:phot}, although we don't
include a separate color evolution figure because of the difficulty of
k-correction to undefined UV filters and an unknown underlying SED
evolution.

For \eu , we k-correct the observed $z$ data to restframe $u$ to
compare to other SLSNe from \citet{2011Natur.474..487Q}, as shown in
Figure~\ref{fig:uband}.  The k-correction \citep{1996PASP..108..190K}
requires the underlying spectral energy distribution (SED) as input.
Since a representative time series of spectra for SLSNe is not known
(nor whether they behave similarly enough to make this feasible),
instead we fit a blackbody model to the photometry
(Section~\ref{blackbody}), and use this as the SED for the
k-correction.  Figure~\ref{fig:uband} shows that SLSNe have diverse
behavior in the $u$-band.  \eu\ reaches a peak magnitude of -22 (AB),
or -22.7 (Vega), comparable to the most luminous SLSNe, although its
rise time is relatively short.  Note that due to the considerable
uncertainties associated with the k-correction for SLSNe, the $u$-band
magnitudes for the PTF objects in Fig.~\ref{fig:uband} are uncertain
by several tenths of a magnitude (Quimby, private communication).
K-corrections for \eu\ are less uncertain, but due to the use of a
blackbody model for the SED, are probably not better than a tenth of a
magnitude.  \eu\ is thus comparable in absolute magnitude to the most
luminous hydrogen-poor SLSN, SN 2005ap ($M=-22.7$), according to
\citet{2012Sci...337..927G}.

\subsection{Blackbody fits\label{blackbody}}
\begin{figure}
\plotone{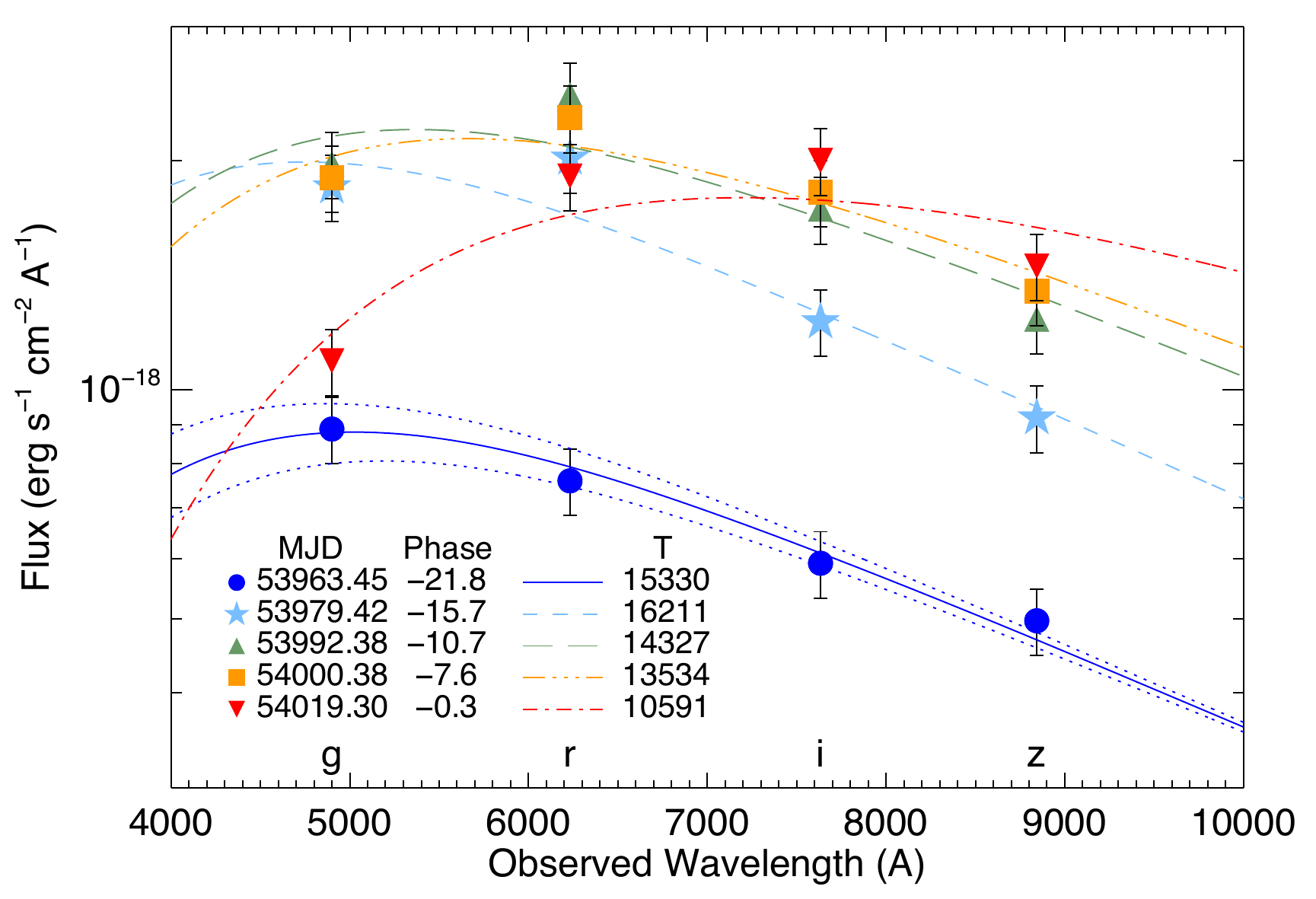}
\caption{Blackbody fits to the photometry for \eu\ on each day where 4
  bands of photometry were available.  Data points are located at the
  central wavelength of each filter.  Only dates where all four bands
  were taken within 24h were considered.  The flux from each day is
  fit independently for the radius of the emitter and the temperature
  of the blackbody.  On restframe day -22 the SN was greater than 15000K
  (solid line), but by maximum light it had cooled to $\approx
  10600$K (dot-dashed line). Since the supernova is not a perfect
    blackbody, Poisson errors underrepresent the true error on the
    blackbody fit.  $\chi^2$/DOF of order unity is obtained with
    errors that are 10\% of the flux (shown).  The temperatures are
    relatively insensitive to the error model.  The dotted line shows
    the effect of varying the temperature by $\pm 500$ K while keeping
    the radius fixed.  It is only shown for the first epoch for
    clarity, though other phases are similar.  Formally, the
    temperature is higher on day -15.7 compared to day -21.8, but this
    is likely because the observations are on the Rayleigh-Jeans tail
    for this phase, and the true temperature cannot be discerned
    accurately.
  \label{fig:bbday2}}
\end{figure}

We can explain the behavior of the lightcurve as an expanding
photosphere radiating as a cooling blackbody.  For \eu , for each day where we
have 4-band ($griz$) photometry taken within 24 hours, we fit a
blackbody to the measured fluxes, as shown in Figure~\ref{fig:bbday2}.
At 22 days before peak brightness the SN is at $T\sim 15000$ K,
decreasing to around 10,600 K near maximum light.  For the radius of
the photosphere at -21.8 days we find $1.7 \times 10^{15}$ cm.
Unfortunately, the SN had been detected for about a week before this
date, and this large radius does not allow us to differentiate between
possible progenitor models.  At maximum light the best-fit radius was
$6.1 \times 10^{15}$ cm.

\begin{figure}
\plotone{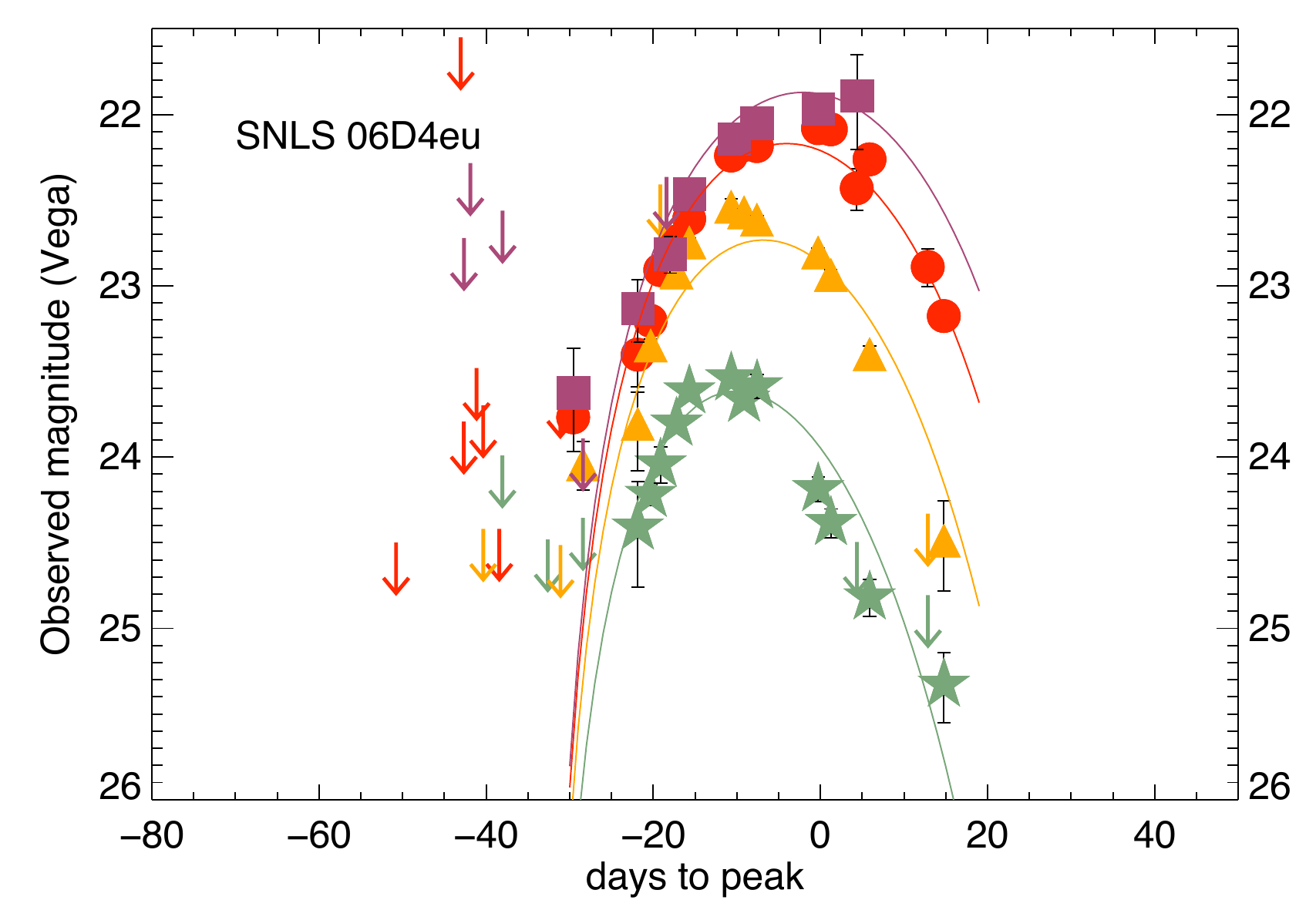}
\caption{Points are photometry of \eu\ as in Fig.~\ref{fig:phot}.
  Solid lines are the simple blackbody model described in
  Sec.~\ref{blackbody}, which has a linearly increasing radius and
  linearly increasing temperature.  Parameters of this model are given
  in Table~\ref{tab:bbparams}.  The time axis is given in the
  restframe.\label{fig:bbfit}}
\end{figure}

To make better use of all the data (not just those epochs with 4-band
photometry), we assume a simple model: 
expansion with the temperature decreasing linearly with time.  The
radius of the photosphere is given by:
$$R(t)=R_0 + \frac{dR}{dt} t,$$
where $R_0$ is the initial radius, $dR/dt$ is the rate of change of
the position of the photosphere with time, and $t$ is time in days.
Meanwhile, the temperature evolution is given by:
$$T(t)=T_0-\frac{dT}{dt} t,$$
where $T$ is the temperature in Kelvin, and $T_0$ is the initial
temperature.  Figure~\ref{fig:bbfit} shows a fit of this simple model
to the data, and the best-fit parameters are given in
Table~\ref{tab:bbparams}.  At day -30, the SN starts with $T\sim
16000$ K and the temperature decreases by about 200K per day.  

Considering the simplicity of the model, the fit to the photometry is
quite remarkable.  It correctly fits the colors of the lightcurves and
the relative timing of the peaks in each band.  In other words, each
lightcurve is not fit independently, they are all fit at once with the
colors set by the model.  The model is a relatively poor fit to the
lightcurve at the earliest times and in the $r$-band.  Of course,
given the relatively strong, broad lines in the spectrum, no blackbody
fit is expected to be perfect.  

The earliest data point in the $r,i,$ and $z$ bands deviates
significantly from the simple model.  This may indicate that the SN
does not follow a simple blackbody at the earliest times.  Some
core-collapse supernovae have an initial decrease in the lightcurves
from the SN shock breaking out of the progenitor star, before they
start to rise.  However, \citet{2012A&A...541A.129L}
saw an unexplained plateau that could not be explained by shock
breakout in the early phase of the lightcurve for the SLSN SN 2006oz.
With only one epoch's data deviating from a very simple model, we do
not speculate further on the cause.

\begin{table}
\begin{center}
  \caption{Best-fit parameters for simple blackbody model. Radii are
    relative.\label{tab:bbparams}}
\begin{tabular}{ll}
\tableline\tableline
$R_0$ & 2.7\\
$\frac{dR}{dt}$ & 1.03 day$^{-1}$\\
$T_0$ & 16270 K\\
$\frac{dT}{dt}$ & 186 K day$^{-1}$\\
\tableline
\end{tabular}
%\tablenotetext{a}{Three sigma upper limit}
\end{center}
\end{table}

\subsection{Energy requirements}
The peak absolute magnitude in restframe $U$ is $M_U=-22.7 \pm 0.1$ (Vega),
making \eu\ among the most UV-luminous supernova ever discovered.

From the blackbody fit we derive an approximate bolometric peak
luminosity of $3.4 \times 10^{44}$ ergs~s$^{-1}$.  Since we can only
sample the ultraviolet part of the spectral energy distribution, there
is considerable uncertainty in using the blackbody assumption to
extrapolate to a bolometric luminosity.  However, the total radiated
energy is approximately $10^{51}$ ergs (1 Bethe [B]), comparable to that of other hydrogen-poor superluminous supernovae \citep[see Table 1 of][]{2012Sci...337..927G}.

\section{Spectra}
\subsection{Observations}
Most of the discussion from here on will focus on \eu , since unlike \sbv , the host redshift is known, thus allowing fits to spectra and the accurate calculation of energetics.

The spectrum of \eu\ was taken at -17 days with respect to $U$
maximum, making it one of the earliest spectra from a superluminous SN
(the time of the spectrum is denoted with an `S' in
Fig.~\ref{fig:phot}).  From our simple expanding blackbody model, we
predict $T=13 900$ K at this epoch.  Figure~\ref{fig:bandpass} shows
that continuum of the observed spectrum roughly matches a blackbody at
$T=13 500$ K (see Fig.~\ref{fig:bbday2} for the effect of changing the
temperateure by $\pm 500$ K).  This is consistent with the blackbody
model and independent verification of the high temperatures reached in
the explosion.

\begin{figure}
\plotone{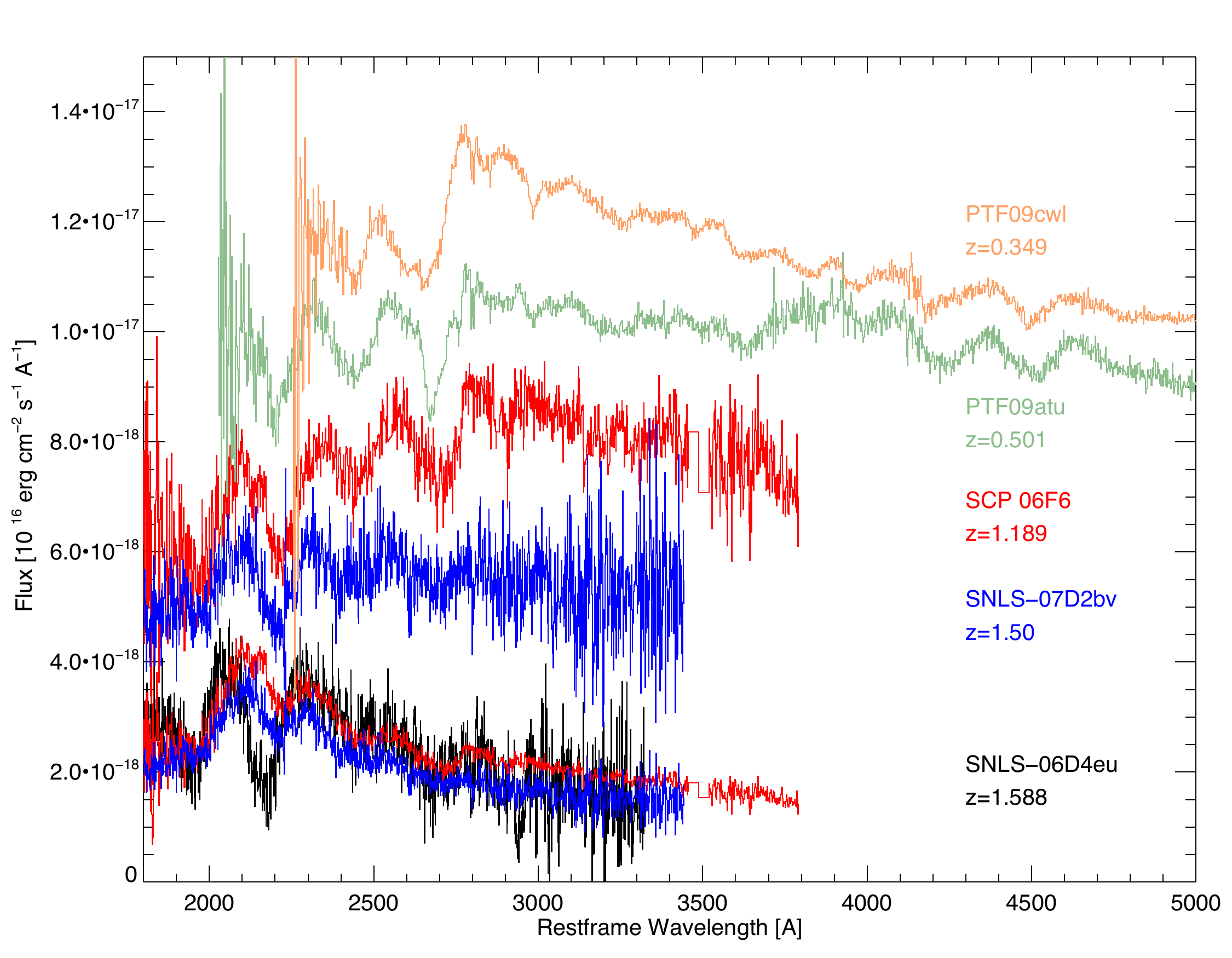}
\caption{The same as Fig.~\ref{fig:spec}, except also showing SCP 06F6
  and SNLS~07D2bv dereddened.  This is not meant to imply
  that they are reddened due to dust.  Instead, they are redder
  because they were observed later in phase than \eu .  SCP 06F6 was
  observed near maximum light in the observer-frame $i$-band
  (Fig.~\ref{fig:phot}).  \eu\ was observed at -17 days relative to
  restframe $U$ maximum, when it was hotter.  Thus the relative
  shallowness of most of the absorption features in \eu\ may only be
  due to the phase of observations.  Note, however, that the 2200 \AA\
  feature is shallower and has a minimum at a longer wavelength in
  SCP~06F6.  SN features tend to move redward with time as the
  photosphere recedes inwards in mass coordinates as the ejecta
  expand.  The difference may be the result of increased abundance of
  that element in the outer layers or a different chemical composition
  between the two supernovae.  Alternatively, there is a hint in the
  spectrum of SCP 06F6 that the line is a blend of two species.  If
  so, higher velocities at early times in \eu\ may more effectively
  blend them.\label{fig:specdered}}
\end{figure}

As Fig.~\ref{fig:spec} shows, the spectrum of \eu\ is quite
different from that of other superluminous supernovae (SLSNe).  It extends farther into the
restframe UV, since it was observed at a higher redshift.  Features
redward of 2200 \AA\ appear depressed in \eu\ relative to the
other events.  \eu\ also peaks farther in the UV.  

The feature near 2200 \AA , seen in SCP~06F6 and PTF09atu is apparent
in \eu , though at a higher velocity.  This feature is
tentatively identified as \ion{C}{2} by \citet{2011Natur.474..487Q}.
In fact, the feature may consist of two separate lines in the former
two supernovae that are blended, possibly because they are formed at a
higher velocity in \eu .  In addition there is an apparent
absorption feature near 2000 \AA\ that was first seen in SCP~06F6, but
is now confirmed in \eu .

At first glance it is not apparent that \eu\ is related to the other
supernovae.  However, it is important to remember that they were
observed at very different epochs relative to maximum light.  Since we
have seen from the lightcurve evolution (Fig.~\ref{fig:phot}) that
these supernovae get dramatically redder as they reach maximum light,
to make an appropriate comparison between a spectrum taken at -17 days
(\eu ) and one taken near maximum in the observed $i$-band, we must
deredden the later spectrum.  We do this using the IDL astrolib
routine CCM\_UNRED.  The result is shown in Fig.~\ref{fig:specdered}.
Though we use a reddening law to ``deredden'' the spectrua of
SCP~06F6 and \sbv , we note that this reddening has nothing to do with dust, it
is a natural consequence of an expanding, cooling blackbody.

From Fig.~\ref{fig:specdered} it is clear that the differences between
SCP~06F6 and \eu\ are largely the result of the differing
phases at which they were taken.  There is an excellent match to the
features near 2000 \AA , 2500 \AA , and 2700 \AA.  The one feature that
does not match as well, that at 2200 \AA , possible \ion{C}{2}, is
shallower and redder in SCP~06F6.  Since the photosphere recedes in
mass coordinates with time in a supernova, and supernova absorption
features generally get redder with time as absorption is happening at
lower velocities, it is not surprising to see this behavior.  In fact,
we might ask why we don't see it in the other absorption lines.  The
relative depth of the 2200 \AA\ feature may be due to differing
abundances, distribution, or ionization state of the element or
elements that give rise to it, possibly \ion{C}{2}. 

Since the lines in these SNe are so far in the ultraviolet, they have
never been seen before in supernovae, making identifications
problematic.  \citet{2011Natur.474..487Q} used SYNOW, a parametrized
code for line identifications that uses the local thermodynamic
equilibrium and other approximations \citep[see Supplementary
Information in][]{2011Natur.474..487Q}.  Here we use a full radiative
transfer model, which should
result in more accurate identifications.

\subsection{Theory}
To further interpret the spectra and constrain the physical properties
of \eu, we ran synthetic spectrum calculations with the Monte Carlo
radiative transfer code SEDONA \citep{2006ApJ...651..366K}.  Rather
than using ejecta models based on first-principles explosion
simulations, we instead attempt to empirically reproduce the spectra
by varying ejecta mass, composition, time since explosion, and total
energy in a parametrized way.

The ejecta density structure was assumed to be spherically symmetric
and to follow a broken power law \citep{1989ApJ...341..867C} with a
shallow inner region ($\rho \propto v^{-1}$) and a steep outer region
($\rho \propto v^{-8}$).  Such a profile is a reasonable approximation
to the density structure seen in hydrodynamical models of core
collapse supernovae.  This model does not, however, capture the
possible existence of global asymmetries or clumpiness, or the
presence of a dense shell due to a magnetar inflated central bubble.

The composition of the ejecta was assumed to be homogeneous.  We
explored five different abundance distributions: (1) solar (2)
helium-rich, in which all hydrogen in the solar composition was
converted into helium, and (3) C/O-rich, in which all the hydrogen and
helium in the solar composition was converted to equal parts carbon
and oxygen, (4) O-Mg-Ne, in which the composition is taken from the
oxygen-magnesium-neon layers of a massive star progenitor from the
stellar evolution models of \cite{2002RvMP...74.1015W}, and (5) IME
burned, in which the intermediate mass elements have been burned,
similar to a Type Ia supernova.  The first four compositions span the
range of possible massive stars depending on how stripped their
envelopes are.

We calculated synthetic spectra at fixed snapshots in time, under the
stationarity approximation \citep{1990sjws.conf..149J}.  Photons were
emitted from the surface of an inner spherical boundary, with a total
luminosity $L$. The emergent spectrum was fairly insensitive to the
location of the inner boundary, as long as it was situated well below
the photosphere; we placed the inner boundary at $v_{\rm core} =
5000$~\kms .  The temperature structure above the photosphere was
calculated self-consistently under the assumption of radiative
equilibrium.

\subsection{Analysis}
\begin{figure}
\plotone{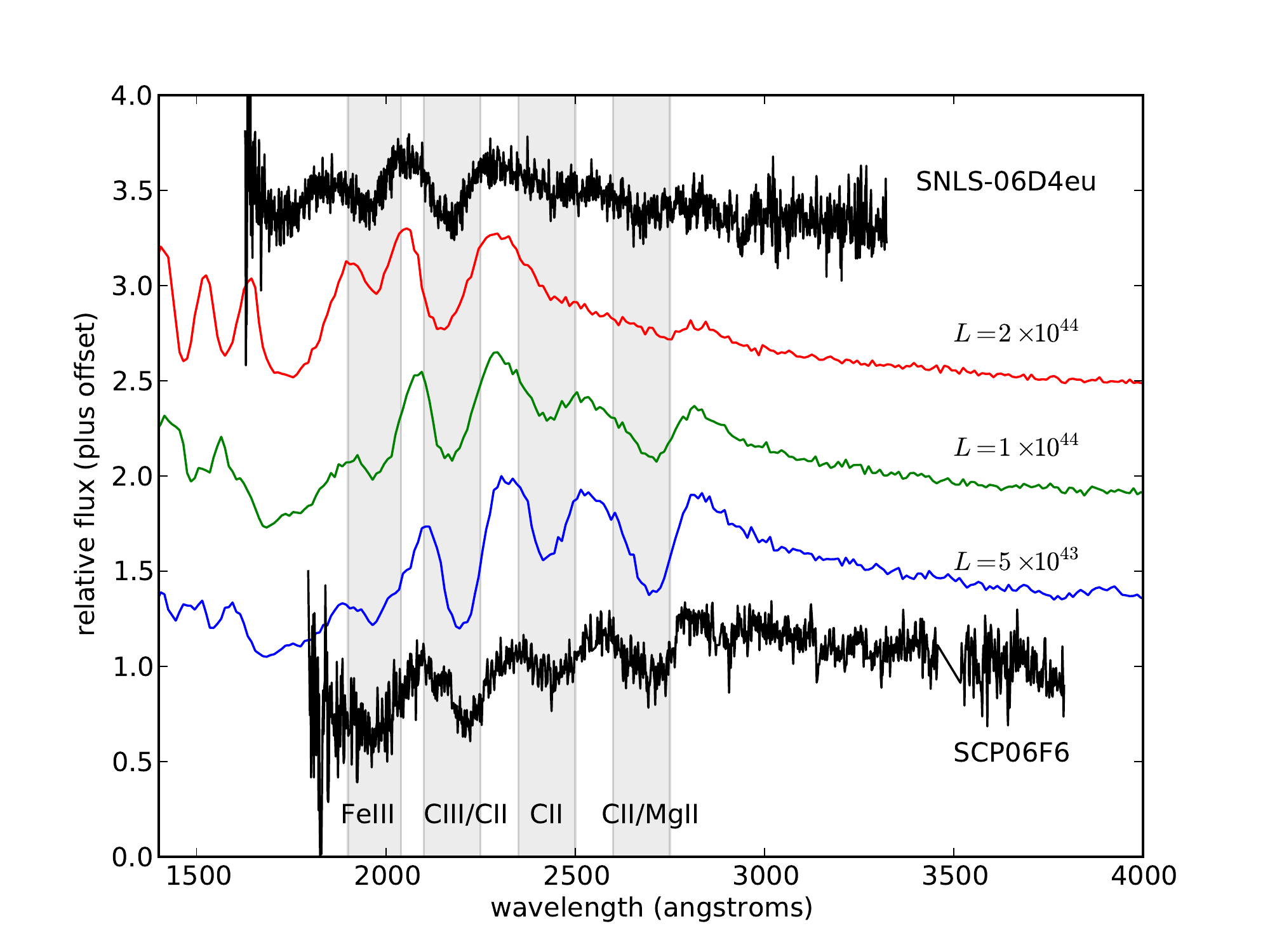}
\caption{{\it Black:} observed spectra of \eu\ at -17d and SCP 06F6
  near maximum light.  {\it Colored lines:} putting different
  luminosities thorough a model using a 5 \Msun\ carbon-oxygen
  progenitor with solar abundances.  Lines of \ion{Fe}{3}, \ion{C}{3},
  \ion{C}{2}, and \ion{Mg}{2} are identified. The data are consistent
  with \eu\ and SCP 06F6 having come from a similar progenitor, but
  having different luminosities, partially due to intrinsic luminosity
  differences, and partially due to them being observed at different
  phases.  Here model spectra
  are computed at a fixed phase (near maximum light), while the
  luminosity was varied. Luminosity is given in ergs s$^{-1}$.\label{fig:specfit}}
\end{figure}

Figure~\ref{fig:specfit} shows the observed spectra of \eu\ and
SCP 06F6 compared to models of varying luminosity, but fixed ejecta
mass ($M = 5$~\Msun ), kinetic energy ($E = 10^{52}$ erg), phase (maximum
light) and a C/O-rich composition.  Probable line identifications are
marked on the figure lines, and are based on determining the lines
with the highest Sobolev optical depth \citep{1960mes..book.....S} near the photosphere.  Firm
line identification is difficult given that most features are blends,
with the relative strength of the different components varying with
luminosity and time.

By using the same mass and composition, but different luminosities for
the models in Figure~\ref{fig:specfit}, we achieve spectra that look
like SCP 06F6 in the least luminous case, and like \eu\ in the most
luminous case.  This is consistent with the idea that these SNe are
physically related, but \eu\ is exceptionally luminous, and was seen
at an earlier phase where the temperature was higher.

The best-match luminosity for \eu\ is $L = 2 \times 10^{44}$~ergs s$^{-1}$,
which corresponds to the approximate bolometric luminosity at the time
the spectrum was observed (17 days before maximum light).  In
subsequent calculations, we fix the luminosity at this value.

Note that in Figure~\ref{fig:specfit}, in the interest of varying only
one parameter, the luminosity was changed, while the phase was fixed
at maximum light.  While this phase is appropriate for SCP 06F6, \eu\
was observed 17 days before maximum light.  Fortunately, the model is
not very sensitive to the assumed time from explosion.  In subsequent
calculations for \eu , we adopt a phase of 25 days since explosion for
the date of the spectrum.  Taken at face value, the simple blackbody
model had a rise time of 30 days, so day -17 corresponds to 13 days
after explosion (Fig.~\ref{fig:bbfit}).  However, the spectral fit is
not quite as good at day 13 after explosion.  In particular, the
\ion{Fe}{3} feature at 2000 \AA\ is not as well fit.  This is because
\ion{Fe}{3} is mostly ionized at these early epochs due to the high
photospheric temperatures.  Since it is clear from the earliest
lightcurve points that the simple blackbody model is a poor
approximation at early times, this date of 13 days since explosion
cannot be taken too seriously.  The observations before -30d are
limits, so the true explosion time is unconstrained.  Many other
supernovae of this class have longer rise times than 30 days, some as
much as 60 days (Figs. \ref{fig:phot} \& \ref{fig:uband}).  Therefore,
we adopt the longer, 42 day rise (making the day -17 spectra
correspond to 25 days after explosion) for all subsequent
spectroscopic models of \eu .

The reasonable fit of the fiducial model Figure~\ref{fig:specfit}
provides strong evidence that the ejecta of \eu\ was rich in carbon
and oxygen.  We identify the feature at 2000 \AA\ as \ion{Fe}{3}, that
at 2200 \AA\ as \ion{C}{3} and \ion{C}{2}, the feature at 2400 \AA\ as
\ion{C}{2}, and the feature at 2700 \AA\ as \ion{Mg}{2} and
\ion{C}{2}.  In contrast, \citet{2011Natur.474..487Q} had identified
the 2200 \AA\ feature as \ion{C}{2}, the 2400 \AA\ feature as
\ion{Si}{3}, and the 2700 \AA\ feature as \ion{Mg}{2}.  They did not
identify a feature at 2000 \AA .  If the feature at 2200 \AA\ is indeed
\ion{C}{3}, then its strength may be a result of the relatively high
temperatures seen at this phase in \eu .

\begin{figure}
\plotone{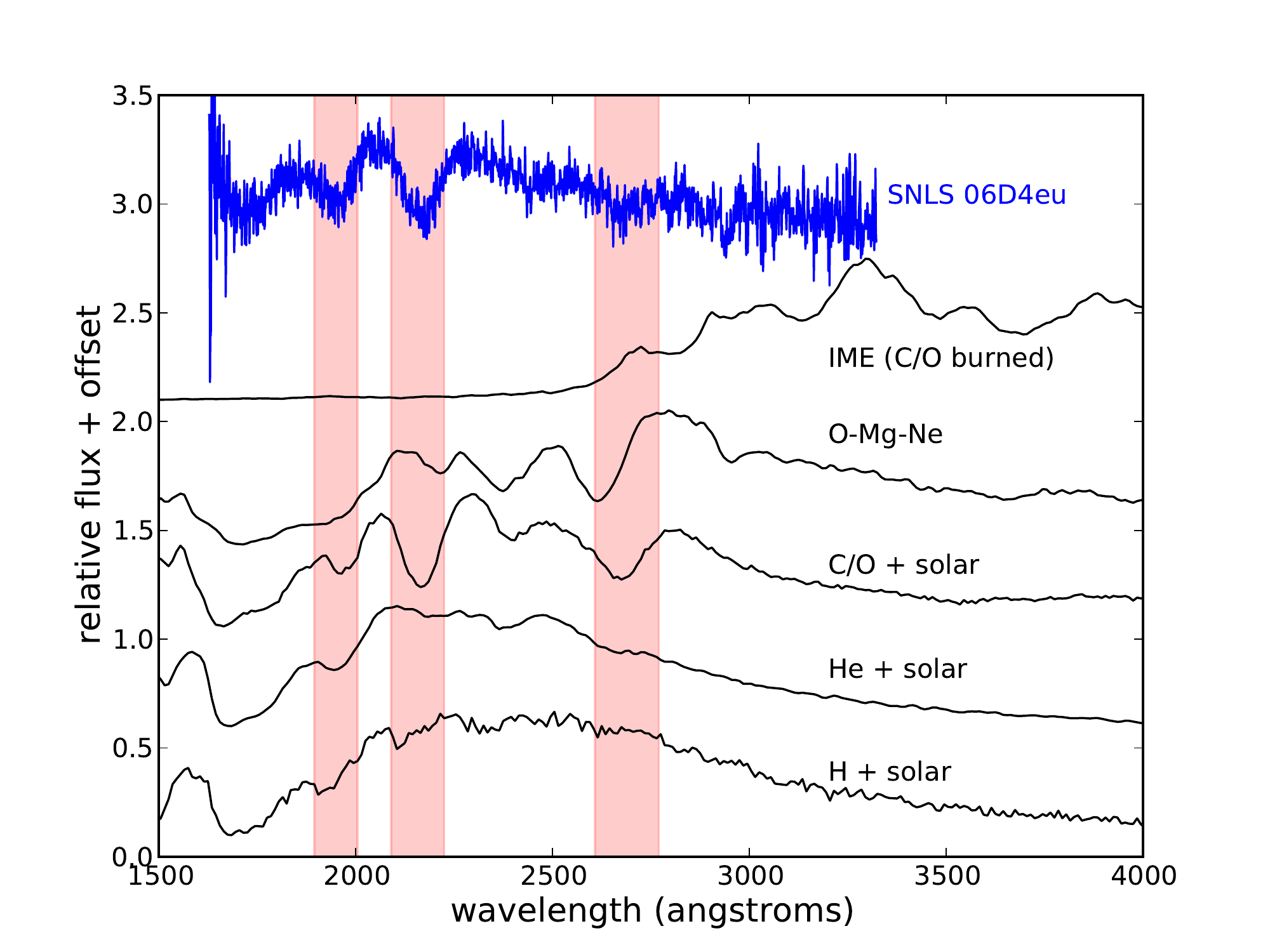}
\caption{Synthetic spectra for models with different ejecta
  abundances.  Each model has an ejecta mass of $M = 5$~\Msun\ and
  kinetic energy of $E = 10$~B (B=Bethe, $10^{51}$ erg).  The spectral
  features of \eu\ are best fit by a model composed of carbon and
  oxygen with a solar abundance of other elements.  
    The IME (C/O
  burned) model is one in which the carbon and oxygen is burned to
  intermediate mass elements, similar to a Type Ia supernova.  
  \label{fig:composition}}
\end{figure}

Figure~\ref{fig:composition} shows that if the composition is instead
assumed to be hydrogen, helium, O-Mg-Ne, or intermediate mass
element rich, none of the major spectral features are reproduced.
When we pair our finding that carbon is required to reproduce the UV
spectra of SLSNe with the previous finding of
\citet{2011Natur.474..487Q} that the most prominent lines in the
optical are \ion{O}{2}, we get a picture of the nature of the SLSN
ejecta as one dominated by C and O.  We therefore associate \eu\ with
the class of Type~Ic supernovae presumed to be the explosions of the
C/O cores of stripped envelope massive stars.  This is consistent with
the finding of \citet{2010ApJ...724L..16P}, who found an empirical
connection between the later-time spectra of SLSNe and SNe Ic.

As seen in Figure~\ref{fig:composition}, the spectra of models with
the H- and He-rich composition are largely featureless, with only a
few broad features that can be attributed to the solar abundance of
metals. No conspicuous lines of H and He are seen, even though these
elements are the dominant constituent of the atmosphere.  This is
because all H and He lines in the observed wavelength range turn out
to be weak given the ionization and excitation conditions in the high
temperature atmosphere.  As a result, we cannot definitively rule out
that the ejecta of \eu\ contained some amount of H and He that is
simply not visible in the spectrum.  However, any layer rich in H or
He would have to exist at high velocities ($\gtrsim 20,000$ \kms),
well above the photospheric layer rich in C-O that is seemingly
required to produce the line features that are seen in the spectrum.

\begin{figure}
\plotone{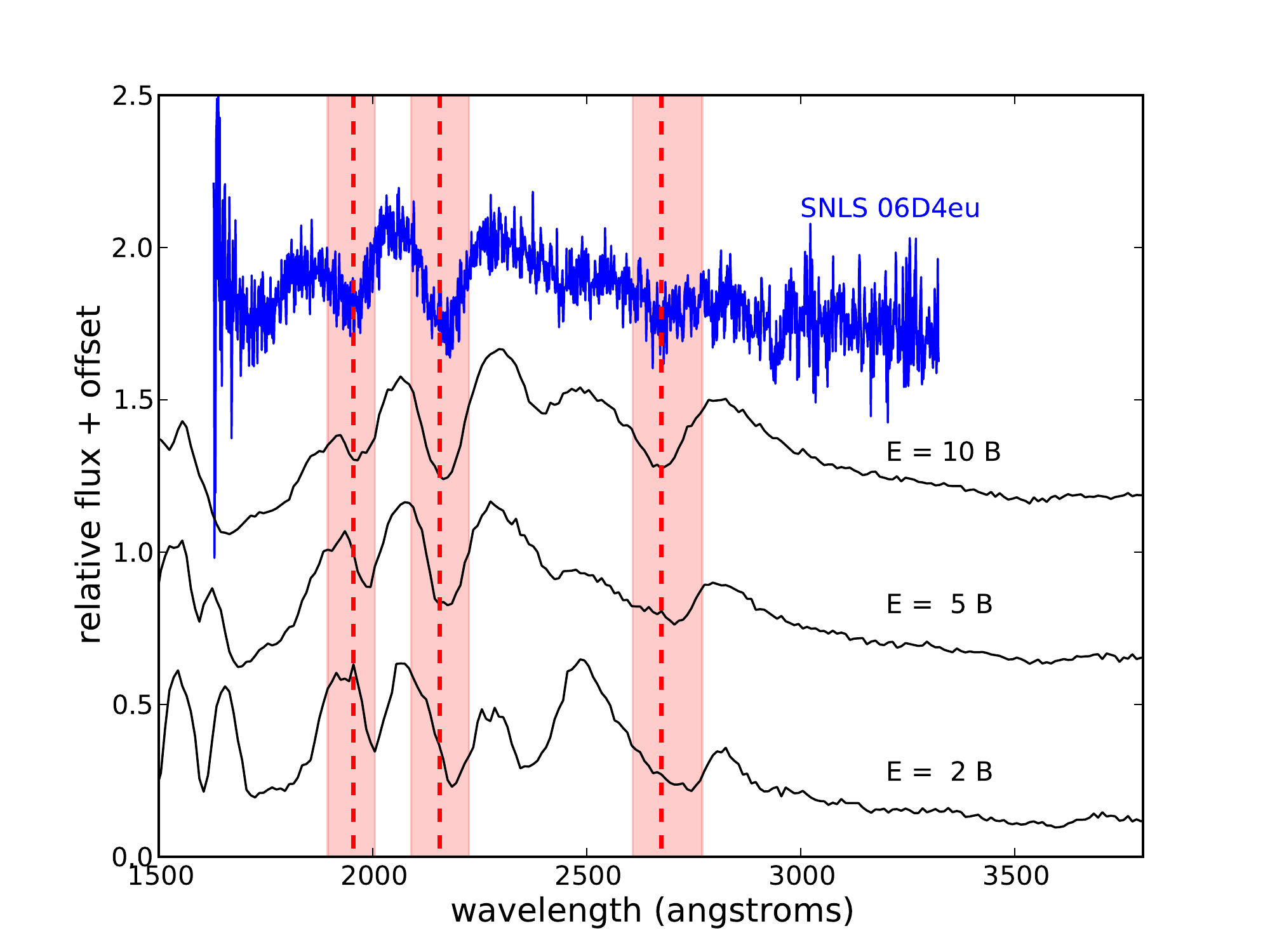}
\caption{Synthetic spectra for models with different kinetic energies.
  Each model has a C/O-rich composition and ejecta mass $M = 5$~\Msun
  .  Dashed lines show the approximate minimum of the absorption
  features in the observed spectrum, while shaded regions show the
  extent of the feature.  The model with $E=10$~B best reproduces the
  blueshifts of the features.  Thus, we require a model with
  relatively high explosion energy, $E/M \sim 2 $B/M$_{\odot}$.\label{fig:energy}}
\end{figure}

It is difficult to constrain the mass and energetics of \eu\ based
only on the spectral fit.  However, the high Doppler shifts suggest a
high kinetic energy.  For $M = 5$~\Msun , we need E = 10 B to reproduce the blueshifts of the features (Fig.\ref{fig:energy}).  Thus it was a
hyperenergetic supernova.  Since the total radiated energy was 1 B,
this also indicates that these SNe radiate energy much more
efficiently than a typical core collapse SN, where the radiated energy
is 1\% of the kinetic energy.

While some studies have speculated that SLSNe result from the
explosion of very massive stars, here a model with only 5 \Msun\ of
ejecta was sufficient to explain the spectra.  While a much higher
ejecta mass ($\sim 100$ \Msun) can not be ruled out based on the
spectrum alone, given the high observed velocities such a model would
require an extremely large energy, $\gtrsim 100 B$. In
Section~\ref{luminosity} we will investigate the mass of the required
progenitor needed to match the observed light curve.

\section{Luminosity\label{luminosity}}
The physical mechanism powering extreme supernovae like \eu\ is
unclear, as the normal core-collapse process does not provide the
requisite $10^{51}$ ergs of radiated energy.  However, there are some
constraints any physical mechanism must meet, including peak output in
the ultraviolet, temperatures in excess of $15000$ K, a 20--50 day rise
and similar decline, and a spectrum devoid of hydrogen, but rich in C,
O, Fe, and Mg.  Here we discuss the hypothesized power sources and
explosion mechanisms.

\subsection{\Ni\ decay}
SNe Ia are powered by the radioactive decay of \Ni .  In the
thermonuclear destruction of a white dwarf, up to a solar mass or more
of \Ni\ is synthesized, and its decay to \Co\ and ultimately \Fe ,
releases gamma-ray photons that are reprocessed and ultimately diffuse
out of the ejecta at UV-optical-IR wavelengths, reaching a peak
luminosity in about 3 weeks \citep{1982ApJ...253..785A}.  However,
\eu\ is more than ten times as bright as a typical SN Ia.
Could it still be powered by the \Ni ?

\begin{figure}
\plotone{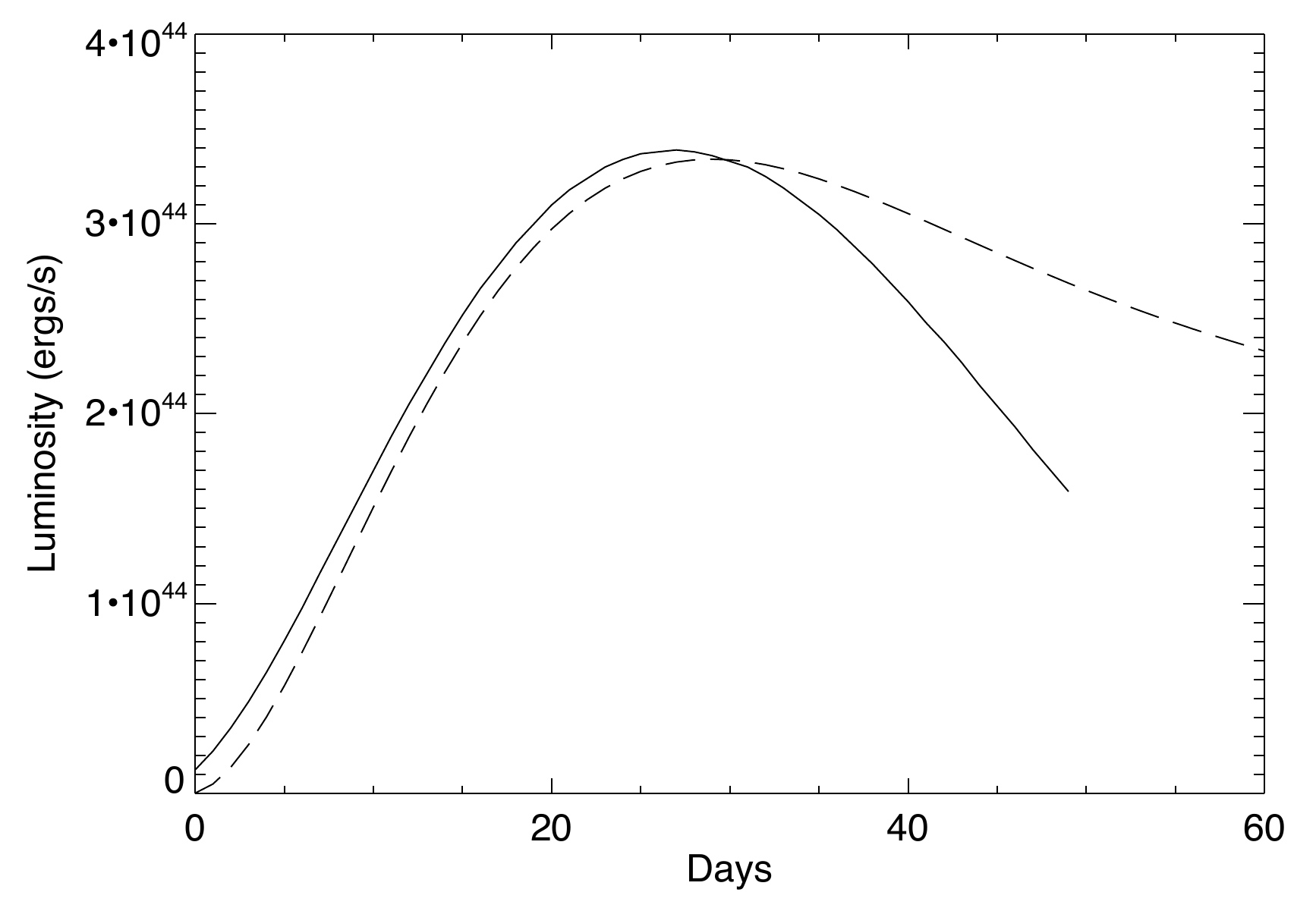}
\caption{Can \Ni\ decay power the lightcurve of \eu ?  The solid line
  shows the bolometric lightcurve of \eu\ derived from the blackbody
  fit.  Days are in the restframe, and day 30 corresponds to the
  fiducial peak in the observed $i$ band of MJD=54020 (day 0 in most
  other plots).  The peak of the blackbody curve is at 26.5d, or 3.5
  restframe days before the fiducial ``day 0.''  The dashed line is
  the simple parametrized model of \citet{2009ApJ...704.1251C} with a
  risetime of 25 days, $v=12000$ km s$^{-1}$ (to match the observed
  spectra), $\kappa=0.1$, $M_{Ni}=23.5$ M$_{\odot}$, and $M_{ej}=25$
  M$_{\odot}$.  A very large mass of \Ni\ is required to explain the
  extreme luminosity, but such a large ejecta mass also requires a
  long diffusion time and slowly declining lightcurve.  No reasonable
  choice of parameters in this model can match the relatively fast
  decline of \eu .\label{fig:nimodel}}
\end{figure}

We use the prescription in \citet{2009ApJ...704.1251C} to calculate a
simple model lightcurve expected form \Ni\ decay
(Fig.~\ref{fig:nimodel}).  A constant opacity of $\kappa=0.1$,
(appropriate for a non-hydrogen atmosphere) is used, as is an ejecta
velocity of 12000 km s$^{-1}$, measured from the spectra.  To fit the
extreme luminosity of \eu , a \Ni\ mass of $M_{Ni}=23.5$ M$_{\odot}$
is required.  Such a large mass of material requires a long time for
photons to diffuse out.  Even for an ejecta mass of $M_{ej}=25$
M$_{\odot}$ (i.e.  only 1.5 \Msun\ of non-radioactive material), the
fall time of the model lightcurve cannot be made to match the more
rapid decline of the bolometric lightcurve.  Furthermore, such a large
ball of \Ni\ , with only a small layer of other material necessary to
reproduce the spectrum, seems implausible.

\subsection{Pair Instability}
Theoretically, very high mass stars ($M \gtrsim 100$ \Msun ), with a core
mass above $\sim 50$\Msun , ought to achieve such high temperatures
that electron-positron pairs are created, decreasing pressure support
in the core, which is followed by core contraction, oxygen ignition, a
large amount of synthesized radioactive nickel, and an explosion that
disrupts the star \citep{1967PhRvL..18..379B}.  This pair instability
mechanism has been invoked to explain the superluminous supernova
SN~2007bi \citep{2009Natur.462..624G}, especially its large
luminosity, slowly declining lightcurve matching $^{56}$Co decay, and
nebular spectra implying a large core and nickel mass.  It may explain
other SNe as well \citep[see][for a review]{2012Sci...337..927G}.
However, \citet{2010ApJ...717L..83M} present an alternative, standard
core-collapse model for SN 2007bi, albeit a massive one ($\sim 100$
\Msun ) that ejects 6.1 \Msun\ of $^{56}$Ni.

Despite its large luminosity, the pair instability model does not seem
to fit \eu .  As indicated in the previous section, the
relatively fast decline of \eu\ argues against a large \Ni\ mass.  The
spectra of \eu\ and other similar objects are also distinct from that
of SN~2007bi.

\subsection{Interaction}
The class of supernovae known as SNe IIn (Type II supernovae with
narrow hydrogen lines in their spectra) can occasionally reach
luminosities in the realm of that seen here
\citep[see][]{2011ApJ...727...15N}.  This is because extra energy is
injected well after the supernova has expanded from its initial radius
by the collision of the ejecta with pre-existing circumstellar
material (CSM).  This interaction produces  relatively narrow
emission lines of hydrogen due to the slow speed of the CSM.  Often
this line is superimposed on a broader line of order hundreds to
thousands of kilometers per second.  Absorption lines are generally
absent, and a featureless blue continuum may dominate at early times.
Furthermore, interaction generally powers the lightcurve for many tens
to hundreds of days, keeping it from showing a strong decline over
this time period.  None of the typical features of SNe IIn match the
observations of \eu\ so we conclude that it is not likely to be
interacting with previously cast-off circumstellar hydrogen from the
progenitor.

Could it be possible that \eu\ and SNe like it are powered by
interaction with something other than hydrogen?  To explain PTF10iue,
a SN Ic with a very slowly declining lightcurve, but no narrow lines
of hydrogen, \citet{PTF10iue} posit that it is a stripped-envelope SN
interacting with a clumpy, hydrogen-free circumstellar medium.  This
would explain the narrow, high-density oxygen emission lines seen in
its spectrum, and its unusually slowly-declining lightcurve, which had
a plateau in the $r$-band for $\sim 200$ days, and declined less than
3 magnitudes over 500 days.  But \eu , \sbv , and similar SLSNe do not
have slowly declining lightcurves nor do they show narrow emission
lines of any element.  It would take an extraordinary degree of fine
tuning to create CSM that was fast-moving enough not to form narrow
lines, was at the right density and morphology to increase the SN
luminosity, but not dilute the spectra, and was perfectly situated to
avoid a plateau in the lightcurve.

\subsection{Pulsational Pair Instability}
\citet{2011Natur.474..487Q} found that one way to achieve the high
luminosities in SLSNe is to invoke an injection of energy at a radius
of about $10^{15}$ cm, possibly though interaction with CSM.  However,
since the SLSNe spectra are dominated by elements such as C, O, Fe,
and Mg, and narrow lines of hydrogen are nowhere to be found, the
normal mechanism powering a SN IIn, would appear not to be at work.
Another possibility is the collision of fast moving shells made of
heavier elements that may be present in the hypothesized pulsational
pair instability supernovae \citep{2007Natur.450..390W}.  These are
theoretically predicted to arise in the deaths of massive stars where
pulsations initially do not succeed in destroying the star, but
instead eject shells of material.  The interaction at a large radius
would overcome the problem of adiabatic losses, the lack of hydrogen
could be explained if there was either previous mass loss or if the
initial pulses rid the star of hydrogen, and the lack of narrow lines
would be explained by the relatively high velocity of ejected
material.  For example, in the 25 $M_{\odot}$ model of
\citet{2007Natur.450..390W}, the initial pulsation produces an
outburst of $\sim 6\times 10^{41}$ erg s$^{-1}$, lasting 200 days.
Subsequent interaction occurs 7 years later, reaching luminosities of
a few $\times 10^{43}$ erg s$^{-1}$.  The final iron core-collapse
then happens after another 9 years.

While the Pulsational Pair Instability supernova model offers
advantages, it requires fine tuning.  As noted in the previous
section, shells must be perfectly placed to add luminosity (without a
plateau) and avoid effects in the spectra.  Such a model would in fact
be different than the initially presented PPI model, which assumed a
star with outer layers of hydrogen.  Further work must be done to
create detailed predictions of lightcurves and spectra from the model
in an attempt to match SLSNe.  One prediction is that precursor
explosions might be observable.  In the 25 $M_{\odot}$ model of
  \citet{2007Natur.450..390W} mentioned above, they would be more than
  6 magnitudes fainter than the peak of \eu.  At $z=1.588$ we have
  little hope of having seen them.  Note, however, a precursor
  outburst was seen 40 days before a supernova in SN 2010mc/PTF10tel
  \citep{2013Natur.494...65O}. It was 4 magnitudes fainter than peak.
  The subsequent supernova was a SN IIn.

\subsection{Creation and Spin-down of a Magnetar}
\begin{figure}
\plotone{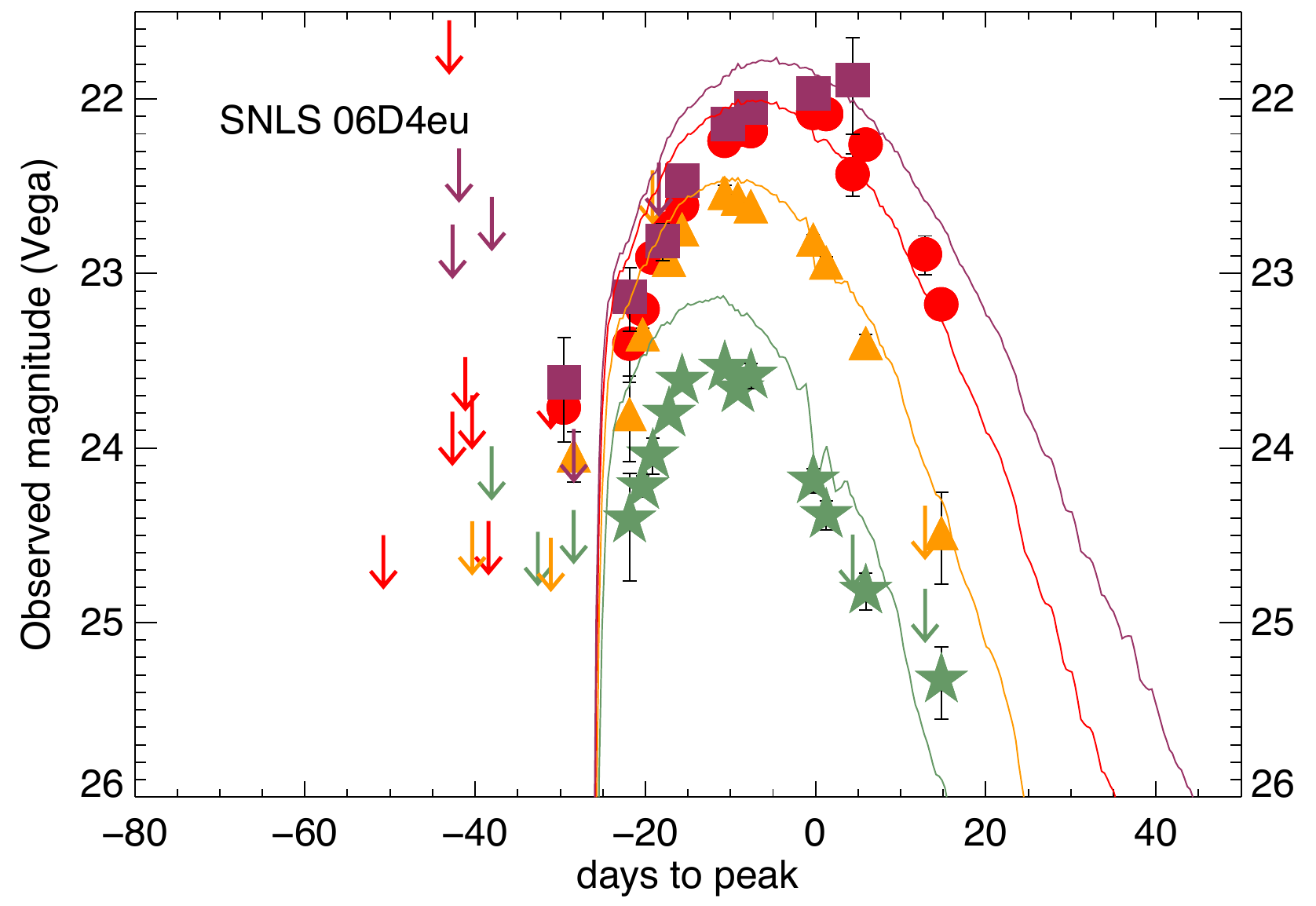}
\caption{06D4eu $griz$ photometry compared to theoretical lightcurves
  from a magnetar spin-down model.  The model was derived from the initial work of \citet{2010ApJ...717..245K}.  This best-fit model has a mass of 3 \Msun , a 
  magnetic field of $B=2 \times 10^{14}$ G and an initial period of
  $P=2$ ms.  A better fit could be achieved with fine-tuning.
  \label{fig:magnetar}}
\end{figure}

\citet{2010ApJ...719L.204W} and \citet{2010ApJ...717..245K}
hypothesize that the creation and spin-down of a magnetar may be able
to inject enough energy at late times into a supernova to account for
the extreme luminosity of SLSNe.  The tunable parameters in this model
are the mass of the progenitor, the size of the magnetic field, and
the starting spin rate of the newly-formed neutron star.  In
Fig.~\ref{fig:magnetar} we compare the lightcurves of \eu\ with such a
model using a 3 \Msun\ progenitor, a magnetic field $2 \times 10^{14}$
G, and an initial period of $P=2$ ms.  From Equation 2 of
  \citet{2010ApJ...717..245K} the spin-down timescale is 24 days.
This was one of a grid of precomputed models without reference to \eu
; fine-tuning could improve the fit.

The model succeeds in matching the overall luminosity, rise and fall
time, and rough flux distribution in $griz$ for \eu .  The match to
the fall time is particularly notable, since supernovae with evidence
for interaction tend to have long plateaus, and models with the
required amount of \Ni\ also have a long fall time.  This is possible
because of the relatively small size of the progenitor, 3 \Msun .
Furthermore, a comparably sized progenitor (5 \Msun ) made of carbon
and oxygen can reproduce the spectra (Fig.~\ref{fig:specfit}) of not
just \eu , but SLSNe in general.

Still, questions remain.  It is also not known how the energy of the
spin-down is coupled to the expanding ejecta.  And these models are
still relatively primitive -- the spectra model and the lightcurve
models are not self-consistent, nor do they contain explosion physics.
They also have several free parameters that can be tuned over a wide
range.  Other possible limitations of the model include that it does
not necessarily predict that the progenitors should be hydrogen-free,
nor does it address the preference for dwarf hosts.  It may be that
more massive stars produce SNe Ic, or that mass loss is associated
with metal poor hosts.

\subsection{Accretion-driven SNe}
Another possibility is an engine-driven supernova, like those thought
to occur in hypernovae.  For example, one can imagine a jet injecting
energy into expanding ejecta at late times, achieving the same result
as that in the magnetar spin-down model.  Since few detailed
models exist, we can only discuss the broadest of implications for
this idea.  \citet{2010ApJ...724L..16P} linked hydrogen-poor SLSNe to
SNe Ic, though the velocities seen here, 12,000 km s$^{-1}$ are not as
high as those seen in the broad-lined SNe Ic linked to GRBs.

Gamma-Ray Bursts do favor low-metallicity, dwarf hosts
\citep{2008AJ....135.1136M}.  But \citet{2010ApJ...721..777A} find
that normal SNe Ic are disfavored in dwarf galaxies, which tend to
produce either SNe Ib or broad-lined SNe Ic instead.  They hypothesize
that metallicity-driven mass loss strips SNe enough to become a normal
SN Ic only in relatively metal-rich environments.  This fits with the
finding that the sites of SNe Ic are more metal-rich than those of SNe
Ib \citep{2011ApJ...731L...4M}.  It is unclear how hydrogen-deficient
SLSNe would fit into this picture.

\citet{2013ApJ...772...30D} consider an accretion-powered scenario, in
which some of the ejecta remains bound, falls back on the newly
created neutron star or black hole, and then drives an outflow (a jet
or disk wind) which re-energizes the outgoing ejecta at relatively
late times ($\sim$weeks).  While they used progenitors that matched those of
SNe Ic, they were unable to match the luminosities seen in \eu .
However, the model is still in its early stages, and the assumptions
about accretion efficiency were relatively conservative.

The accretion model may be observationally difficult to distinguish
from the magnetar model; both involve some central energy source.  The
one main difference may be that the magnetar power likely continues
until late times (with a $t^{-2}$ tail from spin down), while it is
possible that accretion onto the black hole could suddenly cease at
some point (if black hole is powerful enough to blow away all of the
infalling material).  So the black hole model could explain a sharply
falling light curve at later times (without a tail), while this would
be hard to explain with a magnetar.  We do not have the late time data
here to distinguish between the models.  However,
  \citet{2013ApJ...770..128I} do favor the magnetar model for SLSNe
  based on late-time lightcurves.

\section{Conclusions}
As the most distant superluminous supernova with a spectrum
($z=1.588$), \eu\ provides a rare glimpse of the chemical composition
and lightcurve evolution of an early-universe supernova.  It is also
one of the most luminous supernovae known, reaching $M_U=-22.7$ at
peak with a luminosity of $3.4 \times 10^{44}$ ergs s$^{-1}$,
$10^{51}$ erg total radiated energy, and an explosion energy of
$10^{51}$ erg. It is unlike a traditional core-collapse, thermonuclear
or interaction-powered supernova, but is similar to the emerging class
of superluminous supernovae which do not have hydrogen in their
spectra \citep{2012Sci...337..927G}.

\sbv\ is another similar supernova, with $griz$
lightcurves that rise and fall on a similar timescale, and which start
out blue and become redder over time.  While its redshift is not known
with certainty, it is likely at $z\sim 1.5$ based on a match to
\eu .  If so, it is $\sim 1$ mag dimmer than \eu .  Its
spectrum is similar to that of \eu , albeit redder, which is
consistent with it being taken $\sim 1$ week (restframe) later.

Only a handful of SLSNe are known, so their ensemble properties are
not yet fully understood.  Indeed, there is considerable uncertainty
about where to draw the boundaries between SLSNe with different
properties.  The fact that two supernovae show such striking
similarities, and match a subset of other known SLSNe, points to a
common origin.  These two SNe provide the first detailed look into
restframe ultraviolet of these SNe, where they radiate most of their
energy.  In addition, they are the most complete lightcurve of
a SLSN in multiple bands, stretching from well before maximum light
to past maximum.  

The spectra presented here are some of the earliest, bluest spectra of
these events, taken 17 days (06D4eu) and 10 days (07D2bv) before
restframe $U$-band maximum.  They bear similarity to one of the other
SLSNe with a restframe ultraviolet spectrum, PTF09atu, but are
apparently hotter, and are dominated by two strong absorption lines
between 1700 and 2000 \AA\ that we identify as a blend of \ion{C}{3}
and \ion{C}{2}, and \ion{Fe}{3}.  Other lines \ion{C}{2} and a blend
of \ion{C}{2} and \ion{Mg}{2} are weak in \eu , but are more apparent
in the first identified SLSN, SCP 06F6.

According to our model, these lines should be weaker at higher
luminosities, which is consistent with the extreme luminosity of \eu .
By fitting a simple blackbody model to the lightcurve of \eu ,
we find temperatures initially in excess of $15000$ K, dropping by
about 200 K per day.  At the time the spectrum was taken the supernova
was between $13000$ and $14000$ K.  This accounts for some of the 
strange ionization states, such as \ion{C}{3}, rarely seen in supernovae.

The host galaxy redshift of \eu\ of $z=1.588$ was determined from VLT
X-shooter observations taken after the SN had faded.  Using X-shooter
we were able to detect host lines of H$\alpha$, H$\beta$, \ion{O}{2},
\ion{O}{3}, and \ion{N}{2}.  From this we determined that the host of
\eu\ is forming stars at a rate of approximately 8 \Msun\
yr$^{-1}$ and has a metallicity between solar and one-quarter solar.
These findings are consistent with those of
\citet{2011ApJ...727...15N}, who found that similar SNe Ia are
preferentially found in UV-bright hosts that are often low mass and
possibly slightly metal poor.  

We can reproduce the day -17 spectrum of \eu\ with a model with 5
\Msun\ ejecta, total energy $E=10$~B, C/O plus solar composition,
bolometric luminosity $2 \times 10^{44}$ ergs s$^{-1}$, and a phase
approximately 25 days after explosion.  A high luminosity is required
to reproduce the blue spectra and ionization states seen, a low mass
is required to produce the rapid rise and fall of the lightcurve, and
high kinetic energy are required to reproduce the blueshifts of the
features.  When combined with the C/O composition, from a progenitor
perspective, these SLSNe appear to be related to SNe Ic.

The mechanisms powering normal supernovae as presently understood:
radioactive decay, the prompt injection of the energy of gravitational
collapse into a stellar envelope, or the interaction of a supernova
with hydrogen-rich circumstellar material, cannot explain the extreme
energies and lack of hydrogen seen in \eu\ and similar events.
Models invoking a massive generation of \Ni , would require 23.5
\Msun\ to be synthesized, and with so much ejecta, cannot reproduce
the relatively rapid lightcurve decay of \eu .  Therefore, the
pair-instability model invoked for SN~2007bi
\citep{2009Natur.462..624G} would not seem to be viable.

The magnetar spin-down model of \citet{2010ApJ...717..245K} \citep[see
also ][]{2010ApJ...719L.204W}, with an initial period of 2 ms, a
magnetic field of $2 \times 10^{14}$ G, and a progenitor of 3 \Msun\
can account for the total energy required, and can match the general
rise, fall, and color behavior of \eu .  However, this
model offers no particular explanation for the hydrogen-poor nature of
these SNe, or their predominance in high star-forming, possibly metal
poor galaxies.

On the other hand, the pulsational pair instability model
\citep{2007Natur.450..390W} may address the lack of hydrogen, if it
was lost in a previous outburst, and could explain the preference for
high-star-forming, UV-bright environment if a massive progenitor is
required. However, detailed spectroscopic and lightcurve predictions
have yet to be performed for the model.  Engine-driven SNe, where a
jet injects energy at late times, could also achieve the high
luminosities seen in SLSNe, but in the absence of sophisticated
models, this remains speculative.  Alternatively, these supernovae
could be the result of a process not yet imagined.

It is ironic that the brightest supernovae known have been missed by
observers for centuries.  Partly this is because of their rarity: they
may make up only 1 in 10,000 SNe \citep{2013MNRAS.431..912Q}.  But
another reason is that, with their peak output in the UV, Earth-bound
observers have been missing most of the action.  Finally, a third
reason may be their predominance in highly star-forming and possibly
metal-poor galaxies.  These environments would have been common in the
early universe, where the star formation rate was an order of
magnitude higher than it is today, and the build-up of metals had not
yet reached present-day levels.  It is possible that \eu\ is a
relic of an earlier form of supernovae that is all but extinct today.

The future holds great promise for probing these mysterious events in
detail.  Surveys like the intermediate Palomar Transient factory
\citep{2013ATel.4807....1K} and LaSilla / QUEST
\citep{2013PASP..125..683B} are surveying large volumes of sky at low
redshift (the volume surveyed for these luminous events is much
higher), and finding several per year.  But to understand these
supernovae, we need to observe them extensively in the restframe
ultraviolet, where they radiate their peak energies.  For this,
higher-redshift surveys like Pan-STARRS \citep{2011ApJ...743..114C}
and the Dark Energy Survey \citep{2012ApJ...753..152B} are well
suited.

\acknowledgments{D. A. H. acknowledges support from LCOGT.  C. Lidman
  acknowledges the support provided by the Oskar Klein Center at the
  University of Stockholm.  Based on observations obtained with
  MegaPrime/MegaCam, a joint project of CFHT and CEA/DAPNIA, at the
  Canada-France-Hawaii Telescope (CFHT) which is operated by the
  National Research Council (NRC) of Canada, the Institut National des
  Science de l'Univers of the Centre National de la Recherche
  Scientifique (CNRS) of France, and the University of Hawaii.
  Based in part on observations taken at the
  ESO Paranal Observatory (ESO programs 176.A-0589 and 384.D-0222).
  CP and RC acknowledge financial support from the Natural Sciences
  and Engineering Research Council of Canada.  This work was supported
  in part by French state funds managed by the ANR within the
  Investissements d'Avenir program under reference
  ANR--1--IDEX--0005--02.}

{\it Facilities:} \facility{VLT (FORS1)}, \facility{VLT (XSHOOTER)}, \facility{GEMINI (GMOS)}, \facility{CFHT (Megacam)}.


\bibliographystyle{apj}
%\bibliography{astro}

\end{document}